\documentclass[12pt]{iopart}
\expandafter\let\csname equation*\endcsname=\relax
\expandafter\let\csname endequation*\endcsname=\relax
\usepackage{amsmath, amssymb, color, graphicx, stmaryrd, wasysym}
\definecolor{linkcolor}{rgb}{0,0,0.6} 
\usepackage[colorlinks=true,
	pdfstartview = FitV,
	linkcolor    = linkcolor,
	citecolor    = linkcolor,
	urlcolor     = linkcolor,
	hyperindex   = true,
	hyperfigures = false]{hyperref}

\usepackage{amsmath,amssymb,graphicx,cases}
\usepackage{epsfig}
\usepackage{textcomp}
\usepackage{epstopdf}
\usepackage{float}
\usepackage{braket}
\usepackage[normalem]{ulem}
\usepackage{enumerate}
\usepackage{enumitem}
\usepackage{ae,aecompl}
\usepackage{lmodern}

\newcommand{\jbar}{{\bar\jmath}}
\newcommand{\sig}{\sigma}

\usepackage{xcolor,cancel}

\newcommand{\stkout}[1]{\ifmmode\text{\sout{\ensuremath{#1}}}\else\sout{#1}\fi}

\definecolor{mygreen}{rgb}{0.0,0.55,0.3}

\newcommand{\Pe}{\text{Pe}}

\DeclareMathOperator{\arcsinh}{arcsinh}

\begin{document}

\title[Entropy production rate in thermodynamically consistent flocks]{Entropy production rate in thermodynamically consistent flocks}

\author{Tal Agranov}
\address{DAMTP, Centre for Mathematical Sciences, University of Cambridge, Wilberforce Road, Cambridge CB3 0WA, United Kingdom}
\ead{tal.agranov@mail.huji.ac.il}

\author{Robert L. Jack}
\address{DAMTP, Centre for Mathematical Sciences, University of Cambridge,
Wilberforce Road, Cambridge CB3 0WA, United Kingdom}
\address{Yusuf Hamied Department of Chemistry, University of Cambridge, Lensfield Road, Cambridge CB2 1EW, United Kingdom}

\author{Michael E. Cates}
\address{DAMTP, Centre for Mathematical Sciences, University of Cambridge, Wilberforce Road, Cambridge CB3 0WA, United Kingdom}

\author{\'Etienne Fodor}
\address{Department of Physics and Materials Science, University of Luxembourg, L-1511 Luxembourg, Luxembourg}

\begin{abstract}
We study the entropy production rate (EPR) of aligning self-propelled particles which undergo a flocking transition towards a polarized collective motion. In our thermodynamically consistent lattice model, individual self-propulsion is the exclusive source of irreversibility. We derive the fluctuating hydrodynamics for large system sizes using a controlled coarse-graining: our procedure entails an exact correspondence between the EPR evaluated at the hydrodynamic and particle-based levels. We reveal that EPR is maximal when the system adopts a homogeneous configuration, either apolar or polar, and reduced in the non-homogeneous state where a polar band travels in a apolar background {due to strong spatial EPR modulations.} By analyzing the {latter} we also show that asymmetric energetic exchanges occur at the trailing and leading edges, which we map into a thermodynamic cycle in density-polarization space. Finally, we demonstrate that the regime of weak self-propulsion features a singular scaling of EPR, and a non-analyticity of the travelling band profiles.
\end{abstract}

\maketitle


\section{Introduction}

A distinct feature of active matter is the constant injection and dissipation of energy at microscopic scales~\cite{toner_hydrodynamics_2005, marchetti_hydrodynamics_2013}. This dissipation -- and the associated entropy production -- quantifies the departure from equilibrium~\cite{fodor_how_2016,nardini_entropy_2017} and is at the heart of the fascinating collective phenomena in active systems. Studying dissipation has provided insights into transport, response and structural properties~\cite{nardini_entropy_2017, ferretti_out_2024}, and also shed lights on the mechanisms at the basis of some nonequilibrium collective phenomena~\cite{fodor_irreversibility_2022, obyrne_time_2022, ferretti_signatures_2022, yu_energy_2022}. Moreover, conditioning trajectories on non-typical values of dissipation can induce novel collective behaviors~\cite{cagnetta_large_2017, grandpre_entropy_2021, keta_collective_2021, nemoto_optimizing_2019-1, tociu_how_2019, agranov_entropy_2022,agranov_tricritical_2023}.

The framework of stochastic thermodynamics allows quantification of dissipation based on the breakdown of time-reversal symmetry, under the general modelling assumption of local detailed balance~\cite{Katz1984, seifert_stochastic_2012, Maes2021, cates_stochastic_2022}. Applying this framework to active systems is already difficult for a single particle~\cite{ganguly_stochastic_2013, shankar_hidden_2018, dabelow_irreversibility_2019}, and even more challenging for interacting particles ~\cite{gaspard_thermodynamics_2019, markovich_thermodynamics_2021, Boffi_2024, chatzittofi_entropy_2024, bebon_thermodynamics_2024}. {As an alternative modeling perspective}, many-body systems can be {described at a coarse-grained level by field-theories
}~\cite{toner_hydrodynamics_2005, marchetti_hydrodynamics_2013}, in which the \emph{informatic entropy production rate} (IEPR) measures the breakdown of time-reversal symmetry at the hydrodynamic level~\cite{fodor_irreversibility_2022, suchanek2023irreversible, alston2023irreversibility}. In general, {this hydrodynamic IEPR does not match the microscopic dissipation of the individual particles because of the coarse-graining~\cite{fodor_irreversibility_2022, obyrne_time_2022,markovich_thermodynamics_2021,pruessner_field_2022, cocconi_scaling_2022}. It is also insightful to examine the spatial decomposition of EPR~\cite{markovich_thermodynamics_2021, bebon_thermodynamics_2024}, although this procedure is generally ambiguous for field theories~\cite{ro_model-free_2022, nardini_entropy_2017}.}

{For active systems exhibiting motility-induced phase separation (MIPS), EPR has been studied extensively, both microscopically and at the field theoretic level~\cite{fodor_how_2016, nardini_entropy_2017, fodor_irreversibility_2022, obyrne_time_2022, bebon_thermodynamics_2024, chiarantoni_work_2020, ro_model-free_2022}. For one particular lattice based model, it was shown recently that the microscopic dissipation is successfully captured by an exact hydrodynamic field theory~\cite{agranov_entropy_2022,agranov_tricritical_2023}. }
{ For models of aligning particles -- which exhibit flocking~\cite{vicsek_novel_1995,toner_long-range_1995} -- several works~\cite{ferretti_out_2024, ferretti_signatures_2022, yu_energy_2022, borthne_time-reversal_2020-1} have investigated dissipation, and rates of entropy production}. In some of these models~\cite{ferretti_out_2024, ferretti_signatures_2022,  yu_energy_2022}, the dissipation is maximum and singular at the flocking transition. {However, these models either lack a direct link to a microscopic dynamics with a well defined dissipation, or are thermodynamically inconsistent. In particular, dissipation does not vanish at zero self propulsion~\cite{ferretti_out_2024, ferretti_signatures_2022,  yu_energy_2022}.}

As a step towards resolving these difficulties, a key motivation of this work is to analyse the hydrodynamic IEPR while maintaining the connection to microscopic dissipation {in a thermodynamically consistent way}. Specifically, we analyse a model of flocking for self-propelled particles~\cite{agranov_thermodynamically_2024}. The model is thermodynamically consistent since the self-propelled dynamics respects local detailed balance. This property entails that the dynamics reduces to equilibrium in the absence of self-propulsion. Moreover, we derive macroscopic equations of motion for the hydrodyamic fields with an exact coarse-graining. Analyzing the corresponding IEPR, we find -- differently from previous flocking models~\cite{ferretti_out_2024, yu_energy_2022, ferretti_signatures_2022} -- that dissipation does not peak at the ordering transition.

This paper is organized as follows. In Sec.~\ref{model}, we recapitulate the phase behavior of the flocking model~\cite{agranov_thermodynamically_2024}, derive the fluctuating hydrodynamics with sub-leading noise terms, and analyze the weak self-propulsion regime close to equilibrium. In Sec.~\ref{secentro}, we first show that the IEPR of the fluctuating hydrodynamics coincides with the microscopic EPR. Then, we reveal that the EPR is maximal in the homogeneous states (either disordered or ordered), whereas travelling bands are associated with a lower EPR.  For travelling bands, we examine the spatial decomposition of the mean EPR, which we interpreted as thermodynamic cycles where active forces drive currents up and down free-energy gradients. Finally, we study the weak-activity scalings to elucidate the mechanisms sustaining nonequilibrium phase separation. {These results are derived within the framework of a specific model, but we explain that they also provide generic insights.  They also suggest that thermodynamically-consistent flocking models are a relevant platform to study the energetics of nonequilibrium patterns in a broad class of active matter.}


\section{Flocking far from and close to equilibrium}\label{model}

In this Section, we give a brief summary of the phase diagram featuring the flocking transition of our active lattice model and its hydrodynamic description, as already introduced in~\cite{agranov_thermodynamically_2024}. We also present some additional results on the fluctuating hydrodynamics, and the phase behavior close to equilibrium.

\subsection{Lattice model of aligning motile particles}\label{lattice}

\begin{figure}
	\centering
	\includegraphics[width=.48\columnwidth]{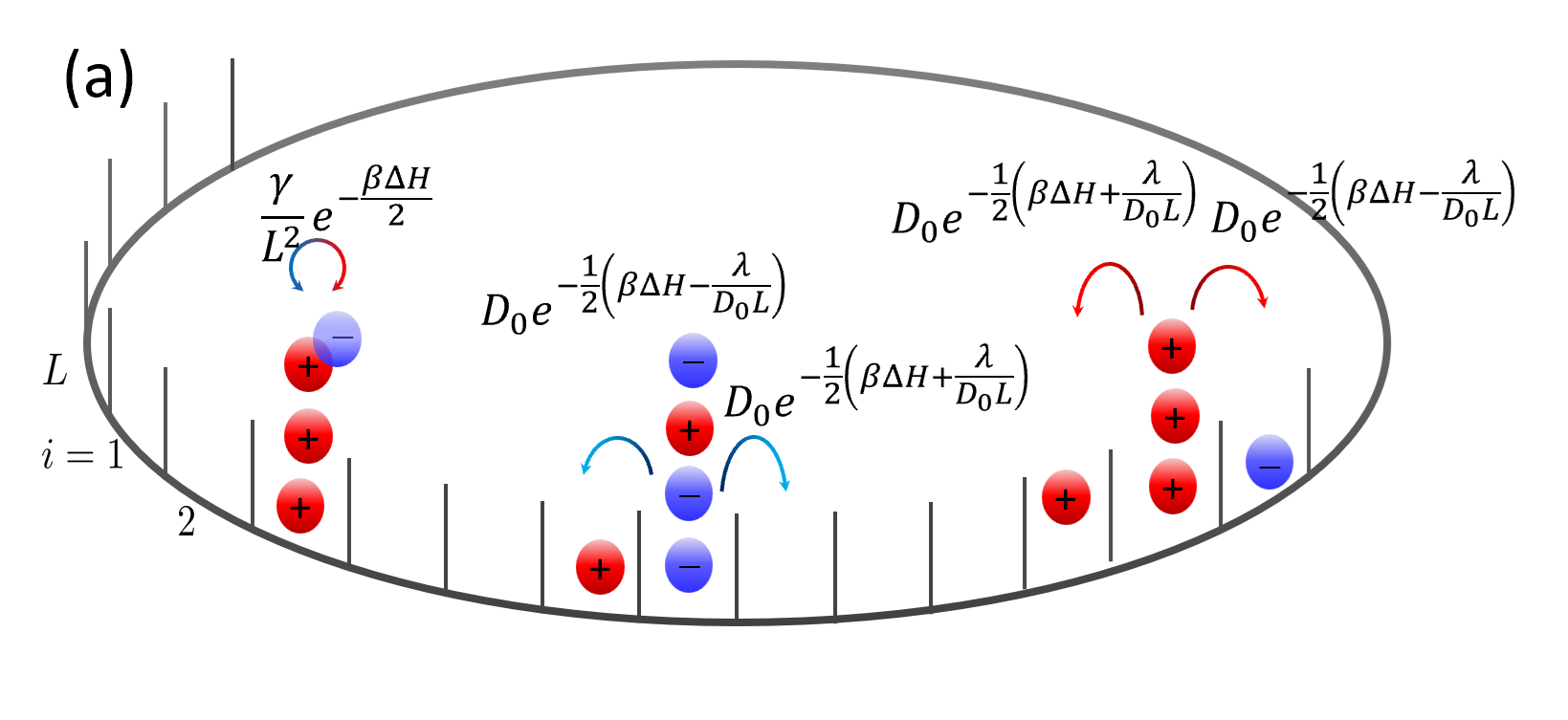}
	\includegraphics[width=.25\columnwidth]{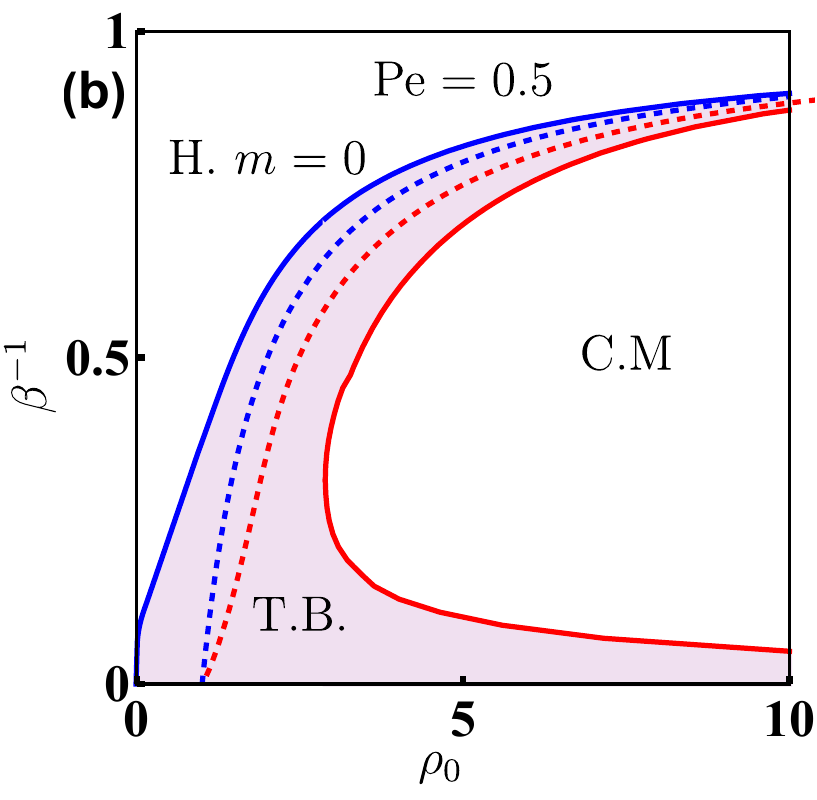}
	\includegraphics[width=.24\columnwidth]{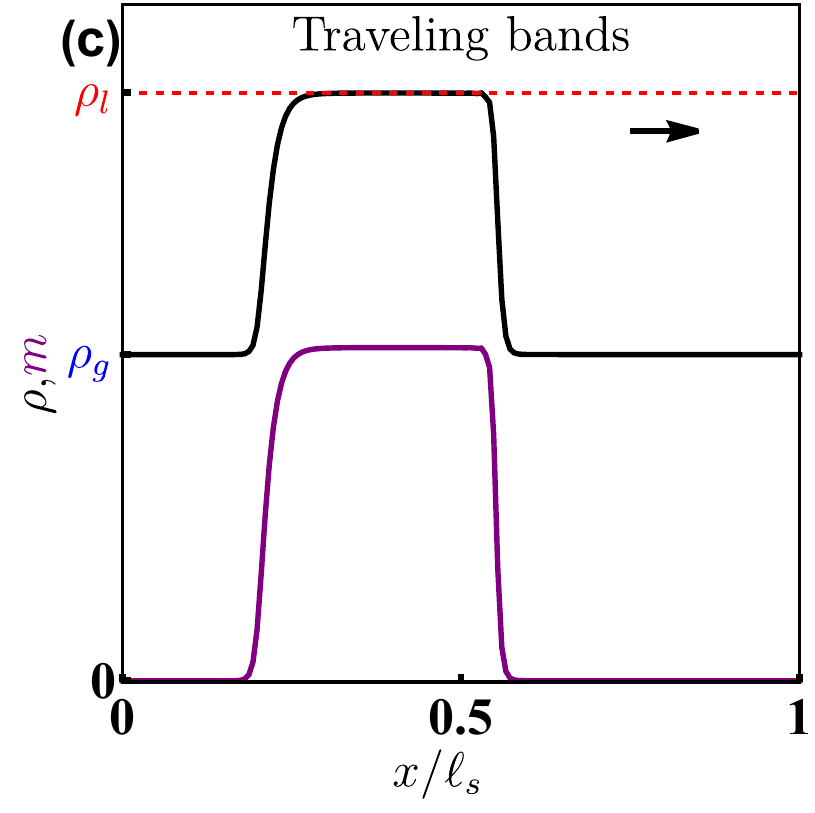}
	\caption{{Summary of results from~\cite{agranov_thermodynamically_2024} for} the lattice model of thermodynamically consistent flocking. (a)~Each particle can be in either one of two states ($+$ and $-$) which determine the direction of biased diffusion (respectively to the right and left). The aligning Hamiltonian $H$ [Eq.~\eqref{h}] constrains both the change of particle states and their diffusive hops. (b)~Phase diagram at finite activity ($\text{Pe}=0.5$). The colored dashed and solid lines represent the spinodals and binodals, respectively. (c)~Profiles of density and magnetization (respectively, $\rho$ and $m$) corresponding to a travelling band, obtained by simulation of the deterministic hydrodynamics [Eq.~\eqref{d2}] with $\text{Pe}=0.5$, $\rho_0\simeq3.62$, and $\beta^{-1}\simeq0.71$. The arrow indicates the direction of band propagation.}
	\label{schem}
\end{figure}

The thermodynamically consistent flocking model comprises, like other discrete symmetry flocks~\cite{solon_flocking_2015, scandolo_active_2023}, two species of particles (type $\sig=\pm1$, although we sometimes write $\sigma=+,-$) which hop with a bias to the right or left (respectively) on a periodic lattice of $L\gg 1$ sites, see Fig.~\ref{schem}. We focus on a one-dimensional lattice but our fluctuating hydrodynamics can be trivially extended to higher dimensions. The dynamics allow multiple occupancy, so that each site $i$ can have any number of particles. The total number of particles in the system is $N=\rho_0 L$, where $\rho_0$ is the mean density.

The dynamical update rules consist of diffusive hops biased by self-propulsion, and spin flips $\sig\to-\sig$. In the absence of self-propulsion, thermodynamic consistency of the model requires that its dynamical rates respect detailed balance with respect to an energy function $H$, whose precise form will be given below. With this choice, the zero-propulsion model has a time-reversal symmetric steady state associated with a vanishing dissipation rate. Previous flocking models do not have this property~\cite{scandolo_active_2023, solon_revisiting_2013, solon_flocking_2015, martin_fluctuation-induced_2021}. Self-propulsion is then added as a weak bias of hops in the direction of the particle orientation, yielding the following update rules [Fig.~\ref{schem}(a)]:
\begin{enumerate}
	\item Site hopping: A particle on site $i$ with orientation $\sigma$ jumps to the neighboring site $i\pm1$  with rate
	\begin{equation} 
	w_\sigma(i\to i\pm1) = D_0 \exp\left[\frac12 \left( \frac{\pm\sigma\lambda}{LD_0} - \beta\Delta H \right) \right] ,
	\label{equ:wi}
	\end{equation}
	where $\Delta H$ is the energy difference between the configurations of the system before and after the jump.\footnote{To obtain the model defined in~\cite{agranov_thermodynamically_2024} one replaces $\sigma\lambda$ in this expression by $2\lambda \Theta(\sigma\lambda)$ where $\Theta$ is the Heaviside (step) function. The two models have identical hydrodynamic behaviour, we choose Eq.~\eqref{equ:wi} here so that the self-propulsion and energetic terms enter on an equal footing.}
	\item Tumbling: A particle changes orientation $\sig\to-\sig$ with rate $(\gamma/L^2)e^{-\frac{\beta\Delta H}{2}}$.
\end{enumerate}
In these rates, $D_0$ is the bare diffusion constant, $\lambda$ is the self-propulsion strength, $\gamma$ sets the tumble rate, and $\beta$ is the inverse temperature. The physical interpretation of Eq.~\eqref{equ:wi} is that the work done by the self-propulsive forces in a single hop from $i$ to $i+1$ is $\lambda\sigma /(\beta LD_0)$: then $w_\sigma$ is consistent with the principle of local detailed balance~\cite{Katz1984, seifert_stochastic_2012, Maes2021}. In particular, provided that the self-propulsion is fueled by some underlying chemical reactions, one can set $\lambda$ as proportional to the difference of chemical potentials for the reactant and product species~\cite{pietzonka_entropy_2018, bebon_thermodynamics_2024}. The $L$-dependence of the rates is standard in lattice-based diffusive models~\cite{kourbane-houssene_exact_2018-1, bodineau_current_2010-1}: it ensures that all processes occur on diffusive time scales in the hydrodynamic limit ($L\to \infty$ at fixed $\rho_0$).

The energy $H$ describes interactions between particles. To obtain exact results at the fluctuating hydrodynamic level, we take these interactions to be long-ranged but weak. Specifically, we introduce the mesoscopic scale $\Delta x = L^\delta$, with $0<\delta<1$, which is sub-extensive in system size while containing a large number of particles (that is, $1\ll\Delta x\ll L$). The interaction range in $H$ is given by $\Delta x$, and it also corresponds to the scale over which we coarse-grain the density to obtain a continuum description in the hydrodynamic limit. In particular, this interaction range enables a phase transition in one dimension even at equilibrium: see~\cite{agranov_thermodynamically_2024} for further details.

To define $H$, we denote by $\eta_i^{+}$ and $\eta_i^-$ the number of $+$ and $-$ particles at site $i$, respectively. We introduce meso-averaged occupancies for the density and magnetization variables in the vicinity of site $i$ as
\begin{equation}\label{coarse}
	\hat{\rho}_i=\frac{\sum_{|i-j|<\Delta x}(\eta^+_j+\eta^-_j)}{2\Delta x} ,
	\quad
	\hat{m}_i=\frac{\sum_{|i-j|<\Delta x}(\eta^+_j-\eta^-_j)}{2\Delta x} .
\end{equation}
We choose the aligning Hamiltonian to take the form
\begin{equation}\label{h}
	H = -\sum_{i=1}^L\frac{\hat{m}_i^2}{2\hat{\rho}_i}g(\hat{\rho}_i) .
\end{equation}
For $g(\rho)=1$, the $i$-th term in the sum is an Ising-like Hamiltonian with interactions of equal strength among all particles within range $\Delta x$ of site $i$. Alternative functional forms for $g(\hat \rho_i)$ modulate the alignment strength according to the local density~\cite{agranov_thermodynamically_2024}.

In the active Ising model (AIM)~\cite{solon_revisiting_2013, solon_flocking_2015}, a non-trivial modulation of the effective interaction strength with density naturally emerges: it can be described as a renormalization effect that captures the effect of fluctuations~\cite{martin_fluctuation-induced_2021}. To qualitatively reproduce this effect, we choose 
\begin{equation}\label{f0}
	g(\rho)=1-\frac{1}{\rho} .
\end{equation}
Other forms of $g$ can be motivated on account of different types of nearest-neighbor interactions~\cite{scandolo_active_2023, gorbonos_pair_2020, bastien_model_2020, king_non-local_2021, wirth_is_2023}.


\subsection{Phase diagram: Apolar gas, polar liquid, and travelling band}
The definition of the model is summarised in Fig.~\ref{schem}(a).  Its phase diagram (for finite self-propulsion) was derived in~\cite{agranov_thermodynamically_2024}, and is shown in Fig.~\ref{schem}(b). It features three phases, which can be characterized in terms of the large-scale behavior of density and magnetization [Eq.~\eqref{coarse}]: (i)~a homogeneous disordered phase (H; $m=0$), where both density and magnetization are homogeneous; (ii)~a homogeneous collective motion (C.M.) phase (H; $m\neq0$), with non vanishing magnetization resulting in a net density flux, and (iii)~a travelling-band state (T.B.) which displays macroscopic phase-separation between different discrete symmetries ($m=0$ and $m\neq0$). In the case (iii), the density and magnetisation vary in space and time as
\begin{equation}\label{eq:travel}
	\rho(x,t) = \bar\rho(x-Vt) , \qquad  m(x,t) = \bar m(x-Vt)
\end{equation}
where $V$ is the velocity of the band and $(\bar\rho,\bar m)$ describe the travelling profile. We refer to the dense and dilute parts of the travelling band as liquid and gas phases, respectively, by analogy with equilibrium phase-separation. The liquid phase is ordered ($m_l\neq0$) while the gas is disordered ($m_g=0$). {The magnetized bulk phase attains a non-vanishing density current, but the disordered phase has none.  Hence particle conservation implies that the interface must propagate with a velocity proportional to the current in the magnetized phase.} Hence, the sign of $V$ is the same as the sign of $m$ in the ordered phase.

The transition between disorder and C.M. is discontinous, and always proceeds through the T.B. state. The same phase behavior is observed for AIM and its recent variants~\cite{solon_revisiting_2013, scandolo_active_2023}. The miscibility gap where T.B. emerges is characterized by two spinodals marking the linear instabilities of the two discrete-symmetry states ($m=0$ and $m\neq0$), whose analytical derivation can be found in~\cite{agranov_thermodynamically_2024}. In the absence of a free-energy minimization principle, there is no simple recipe to establish the binodals in the nonequilibrium case. These are identified numerically from the density profiles [Fig.~\ref{schem}(c)].

Our previous study~\cite{agranov_thermodynamically_2024} explored some different types of phase diagram that emerge if $g$ in Eq.~\eqref{f0} is replaced by a non-monotonic function. A particular choice can lead to a phase diagram with an azeotropic point and some reversed bands, reminiscent of the states reported in some other hydrodynamic studies~\cite{Nesbitt_Uncovering_2021, Thibault_Diversity_2022}. However, we restrict here to the simplest choice in Eq.~\eqref{f0}, which is sufficient to reveal the relevant behaviour of the IEPR.


\subsection{Deterministic hydrodynamics for density and magnetization}\label{hyd}

We present the deterministic hydrodynamic equations that describe the thermodynamically-consistent flocking model on macroscopic length scales. Their derivation follows~\cite{agranov_thermodynamically_2024}. We define the typical length scale for diffusive motion $\xi_d$, the persistence length due to self-propulsion $\xi_p$, and {their ratio, }the Peclet number Pe{,} quantifying the relative contributions of self-propulsion and diffusion:
\begin{equation}
	\label{equ:xi-xi-pe}
	\xi_d = L\sqrt{D_0/\gamma} ,
	\quad
	\xi_p = \lambda L/\gamma,
	\quad
	\text{Pe} = \xi_p/\xi_d = \lambda/\sqrt{\gamma D_0} ,
\end{equation}
We perform spatial coarse-graining introducing the scaled position $x=i/\xi_d\in[0,\ell_s)$, so that 
\begin{equation}
	\ell_s=L/\xi_d=\sqrt{\gamma/D_0}
	\label{eq:ell_s}
\end{equation}
is the system size in the hydrodynamic representation. Within this choice, the interfacial widths of phase separated profiles are of order $1$, and we take $\ell_s\gg1$ so that the system is large compared to this interfacial width. In addition, we make a diffusive scaling of time by setting 
\begin{equation}
	t=\frac{\gamma}{L^2} \hat{t} ,
\label{equ:t-hydro}
\end{equation} 
with $\hat{t}$ the time variable of the microscopic system. Hence, in these hydrodynamic units, particles flip their orientations with unit rate.

The local density in the vicinity of point $x$ is defined from the mesoscopic coarse-graining [Eq.~\eqref{coarse}] as
\begin{equation} \label{den}
	\rho(x,t)=\hat{\rho}_{\frac{xL}{\ell_s}}(t) ,
	\qquad
	m(x,t)=\hat{m}_{\frac{xL}{\ell_s}}(t) .
\end{equation}
The fields $(\rho,m)$ obey some deterministic equations in the hydrodynamic limit ($L\to\infty$). Since $\rho$ is conserved and $m$ is not, these equations take the generic form
\begin{equation}\label{d1}
	\begin{bmatrix}
	\partial_t{\rho}\\
	\partial_tm
	\end{bmatrix}=-\partial_x\begin{bmatrix}
	J_{\rho}\\
	J_m
	\end{bmatrix}
	-\begin{bmatrix}
	0\\
	2K
	\end{bmatrix} ,
\end{equation} 
where $(J_{\rho}, J_m)$ are the conservative fluxes, and $K$ is the net rate of change of the density of $-$ particles due to spin flipping. The definition of hydrodynamic fields in Eq.~\eqref{den} implies that $\rho(x,t)$ is the {locally averaged} occupancy of the site at position $x$, and the lattice spacing is $\ell_s/L$: then, the number of particles in $[x,x+h]$ is $(L/\ell_s)\int_{x}^{x+h} dx \rho(x)$. It follows from Eq.~\eqref{d1} that the (net) number of particles passing to the right through point $x$ is $(L/\ell_s)J_\rho(x,t)$ per unit time. For travelling band states, the velocity $V$ in Eq.~\eqref{eq:travel} reads
\begin{equation}\label{eq:VV}
 V=\text{Pe}\frac{m_l}{\rho_l-\rho_g} ,
\end{equation}
as a consequence of mass conservation.

The fluxes $({J}_{\rho},{J}_{m}, K)$ fluctuate around their most likely values, denoted by $(\bar{J}_{\rho},\bar{J}_{m},\bar K)$.\footnote{Note, the overbars on $\bar{J}_{\rho,m},\bar K$ indicate most likely values of these quantities, dependent on $(\rho,m)$. This notation is distinct from overbars on $(\bar\rho,\bar m)$ in Eq.~\eqref{eq:travel} which are specific to the steady state.} To leading order in system size, these read 
\begin{equation}
\begin{aligned}\label{d2}
	\begin{bmatrix}
	\bar{J}_{\rho}\\
	\bar{J}_m
	\end{bmatrix}
	& =-\frac{1}{2}\mathbb C(\rho,m)\partial_x\begin{bmatrix}
	\frac{\partial {\cal F}}{\partial\rho}\\
	\frac{\partial {\cal F}}{\partial m}
	\end{bmatrix}+\text{Pe}\begin{bmatrix}
	m\\
	\rho
	\end{bmatrix},
	\\
	\bar{K}& =M(\rho,m) \sinh\!\left(\frac{\partial {\cal F}}{\partial m}\right) ,
\end{aligned}
\end{equation} 
where $\mathbb C$ and $M$ are mobility coefficients given by 
\begin{equation}\label{mob}
	\mathbb C(\rho,m)=2\begin{bmatrix} \rho&m
	\\
	m&\rho\end{bmatrix},
	\quad
	M(\rho,m)=\sqrt{\rho^2-m^2} .
\end{equation}
The free-energy functional is defined in units of the temperature ($\beta^{-1}$): 
\begin{equation} \label{eq:FF}
F[\rho,m]=\frac{L}{\ell_s}\int_0^{\ell_s}\!dx\, \mathcal F(\rho(x),m(x))
\end{equation}
with free-energy density
\begin{equation}\label{free}
	\mathcal F (\rho,m)= \frac{\rho}{2}\ln\frac{\rho^2-m^2}{4}+\frac{m}{2}\ln\frac{\rho+m}{\rho-m}-\rho-\beta \frac{m^2}{2\rho}g\left(\rho\right) ,
\end{equation}	
so that $\delta F/\delta \rho = (L/\ell_s) (\partial{\cal F}/\partial\rho)$, and similarly for $m$ derivatives.

The hydrodynamics in Eqs.~(\ref{d1}-\ref{d2}) describes a broad class of thermodynamically consistent active models~\cite{agranov_thermodynamically_2024}. For the particular choice of interactions in Eq.~\eqref{f0}, the dynamics become 
\begin{equation}
\begin{aligned}\label{d1s}
	\partial_t{\rho} & = \partial_x \left[ \partial_x\rho+\beta\left(\frac{m^2}{\rho^3}\partial_x\rho-\frac{m}{\rho^2}\partial_xm\right)-\text{Pe}\,m \right] ,
	\\
	\partial_t{m} & = \partial_x \Big[
	\partial_xm  + \beta \Big( A(\rho,m) \partial_x\rho+ B(\rho,m) \partial_xm\Big) -\text{Pe}\,\rho 
	\Big]
	\\ & \quad - 2m\cosh\left[{\beta m}(\rho-1)/\rho^2 \right] 	+2\rho\sinh\left[{\beta m}(\rho-1)/\rho^2 \right],
\end{aligned} 
\end{equation}
where
\begin{equation}
	A(\rho,m) = m\left(\rho^3-2\rho^2+3m^2-m^2\rho\right)/\rho^4 ,
	\quad
	B(\rho,m) = (m^2\rho-2m^2+\rho^2-\rho^3)/\rho^3 .
\end{equation}
The diffusive fluxes in $\rho$ and $m$ include (i)~a linear diffusion term, stemming from the entropic contribution in $\cal F$ [Eq.~\eqref{free}], (ii)~some non-linear diffusion terms proportional to $\beta$, stemming from the energetic contribution in $\cal F$ [Eq.~\eqref{free}], and (iii)~a self-propulsion term proportional to $\Pe$. The separation of these contributions will prove useful when formulating a decomposition of the IEPR [Sec.~\ref{sec:epr-hydro}]. In particular, the equilibrium hydrodynamic model simply follows by setting $\Pe=0$, and corresponds to a non-ideal reaction-diffusion system~\cite{aslyamov_nonideal_2023, proesmans}.


\subsection{Path-probability representation of hydrodynamic fluctuations}\label{fluc}

The deterministic hydrodynamics [Eqs.~(\ref{d1}-\ref{d2})] has been previously derived in~\cite{agranov_thermodynamically_2024}. We now go beyond by also deriving the sub-leading noise terms that complement these equations. For passive lattice gasses, such corrections are well known within Macroscopic Fluctuation Theory (MFT)~\cite{bertini_macroscopic_2015-1}. Extensions to active gasses appeared recently for related models~\cite{agranov_exact_2021, agranov_macroscopic_2022}. For thermodynamically consistent dynamics, the exact form of the noise can be anticipated from the deterministic description [Eqs.~(\ref{d1}-\ref{d2})], as we discuss below.

In general, hydrodynamic fluctuations can be written in terms of the probability of a given trajectory $\cal X$ of the hydrodynamic fields $(\rho,m)$ and the corresponding fluxes $(J_\rho,J_m,K)$, which we denote by
\begin{equation}
	{\cal X} = \left\{\rho(x,t),m(x,t),J_{\rho}(x,t),J_{m}(x,t),K(x,t)\right\}_{x\in[0,\ell_s],t\in[0,T]} ,
\label{pathX}
\end{equation}
where $T$ is the trajectory duration measured on the hydrodynamic scale, not to be confused with the temperature $\beta^{-1}$. For any trajectory, the fluxes $(J_\rho,J_m,K)$ always obey Eq.~\eqref{d1}, but they differ in general from the deterministic values in Eq.~\eqref{d2}. The probability of such a trajectory $P$ is given in terms of the path action $\cal A$ as
\begin{equation}
	P({\cal X}) \simeq {\rm e}^{-\frac{L}{\ell_s}{\cal A}({\cal X})} .
\label{equ:path-LDP}
\end{equation}
We do not consider any contributions from the initial condition of the trajectory, which are negligible in steady states. Since the hopping and flipping moves are independent [Sec.~\ref{lattice}], the fluctuations of $(J_\rho,J_m)$ and $K$ are uncorrelated. Then, the path action can be expressed as an integral over two additive Lagrangian densities $(\mathcal L_J,\mathcal L_K)$, that respectively embody the diffusive and tumbling fluctuations:
\begin{equation}
	\mathcal{A}= \int_0^T dt\int_0^{\ell_s} dx\, (\mathcal L_J+\mathcal L_K).
	\label{act}
\end{equation}
Interestingly, the expression of $(\mathcal L_J,\mathcal L_K)$ can be directly deduced from the knowledge of the free energy [Eq.~\eqref{free}] and the mobility coefficients [Eq.~\eqref{mob}]. We summarise the results here, see~\ref{ap.fluc} for a more detailed discussion.

The diffusive scaling of space and time [Sec.~\ref{hyd}] ensures that the fluctuations of $(J_\rho,J_m)$ are Gaussian at the hydrodynamic level~\cite{bertini_macroscopic_2015-1}: it results in a Lagrangian $\mathcal L_J$ which is quadratic in the deviations from the flux average values $(\bar J_\rho,\bar J_m)$. In equilibrium ($\text{Pe}=0$), local detailed balance prescribes that the corresponding quadratic form reads
\begin{equation}\label{fluxrate}
 \mathcal L_J { (J_\rho,J_m,\rho,m) } = \frac{1}{2}
\begin{bmatrix}
	J_{\rho}-\bar{J}_{\rho}\\
	J_m-\bar{J}_m
	\end{bmatrix}^\dag
	\mathbb{C}^{-1}	
	\begin{bmatrix}
	J_{\rho}-\bar{J}_{\rho}\\
	J_m-\bar{J}_m
	\end{bmatrix} ,
\end{equation}
where the mean fluxes $(\bar{J}_\rho,\bar J_m)$ and the mobility matrix $\mathbb{C}$ depend on the fields $(\rho,m)$ [Eqs.~\eqref{d2} and~\eqref{mob}]. The expression in Eq.~\eqref{fluxrate} also holds for finite $\text{Pe}>0$, since the self propulsion enters only as a weak bias in the microscopic dynamics [Sec.~\ref{lattice}]: this scaling ensures that local detailed balance is maintained to leading order, as a general result in diffusive lattice gases~\cite{bodineau_current_2010-1}.

The tumbling rates are much slower then the fast diffusive swaps [Sec.~\ref{lattice}]. As a result, the corresponding fluctuations remain non-Gaussian at hydrodynamic scales even after diffusive scaling~\cite{bodineau_current_2010-1}. In practice, the corresponding Lagrangian $\mathcal L_K$ can be expressed as the difference between the large deviation functionals of two Poisson processes associated with tumbling ($\pm\to\mp$)~\cite{bodineau_current_2010-1, kaiser_canonical_2018, kraaij_deriving_2018}. The resulting expression, which is compatible with the detailed balance constraint, is given by 
\begin{equation}
	\mathcal L_K(K,\rho,m)=\sqrt{M^2+\bar{K}^2}-\sqrt{M^2+K^2}+K\left[\arcsinh \frac{K}{M}-\arcsinh \frac{\bar{K}}{M}\right]  .
	\label{phi12} 
\end{equation}
As for the diffusive Lagrangian ${\cal L}_J$, the mean flip rate $\bar{K}$ and the mobility $M$ explicitly depend on the fields $(\rho,m)$ [Eqs.~\eqref{d2} and~\eqref{mob}].


\begin{figure}
	\centering
	\includegraphics[width=.31\columnwidth]{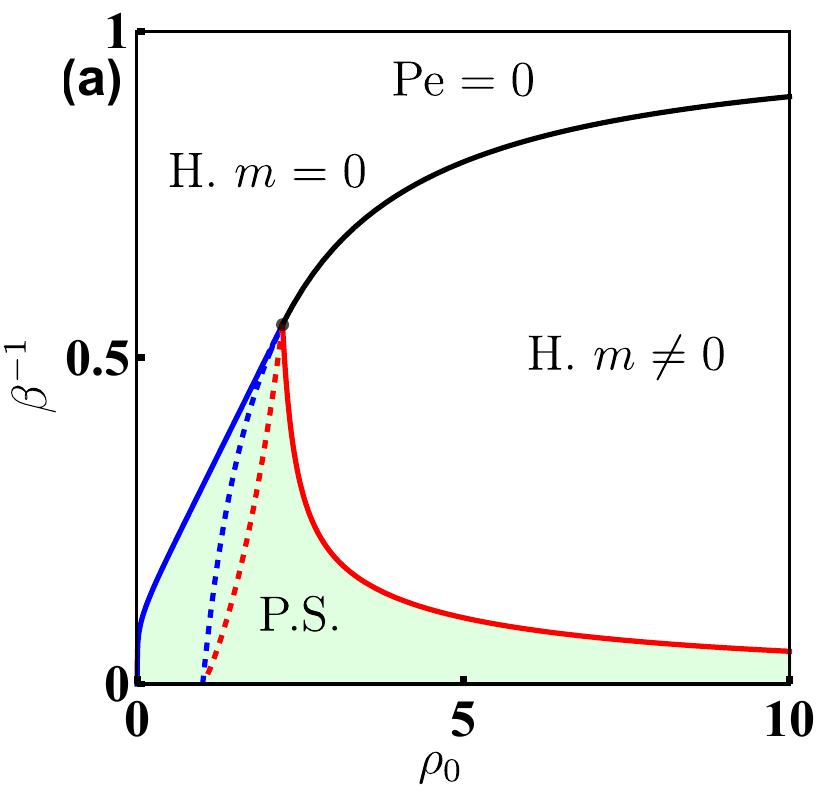}
	\includegraphics[width=.31\columnwidth]{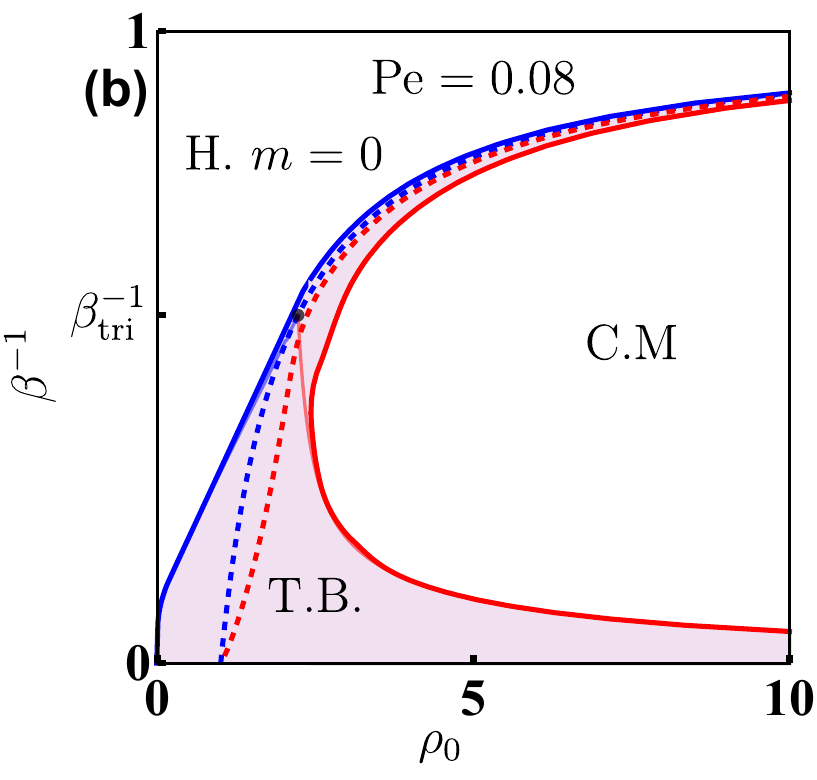}
	\includegraphics[width=.32\columnwidth]{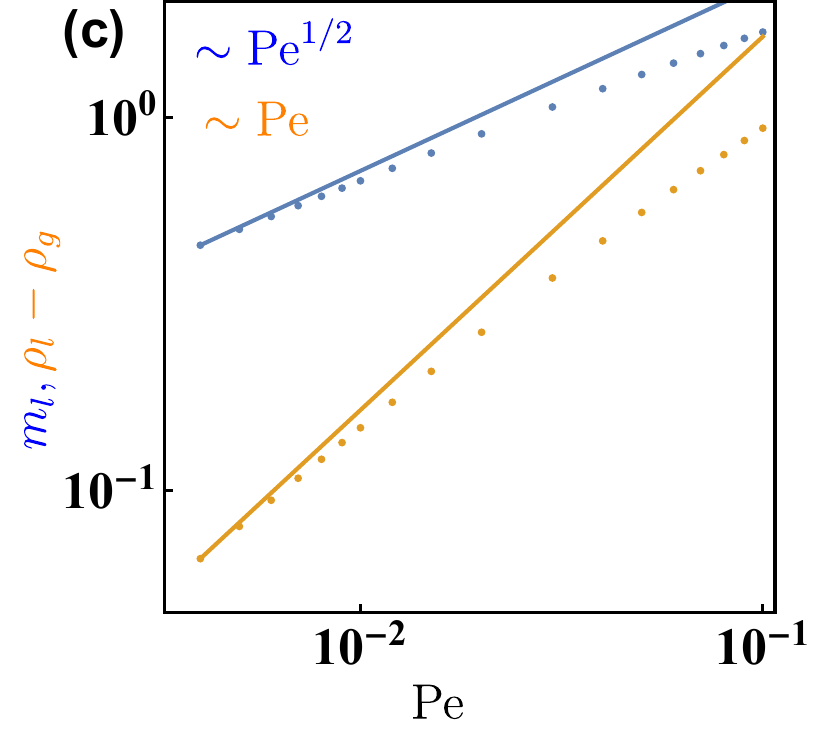}	
	\caption{Phase diagrams (a)~at equilibrium ($\text{Pe}=0$) and (b)~close to equilibrium ($\text{Pe}=0.08$). The faded blue and red curves in panel (b) represent the binodals of the equilibrium limit in panel (a), terminating at the tricritical point ($\beta^{-1}=\beta^{-1}_{\text{tri}}$). In panel (b), the binodal gap extends above the tricritical temperature ($\beta^{-1}>\beta^{-1}_{\text{tri}}$). 
	(c)~The binodal gap $\rho_l-\rho_g$ and the liquid magnetization $m_{l}$, evaluated from the numerical solution of the deterministic hydrodynamics [Eq.~\eqref{d2}] at varying $\text{Pe}$ and for $\beta^{-1}=0.8$. The blue and orange lines have slopes $1/2$ and $1$, respectively, following the analytical predictions [Eq.~\eqref{expand0}].}
	\label{narrow}
\end{figure}

\subsection{Phase behavior close to equilibrium}\label{lowpelimit}

Some of the results of this work analyse the IEPR for small Pe. To facilitate this analysis, we derive here some properties of the phase diagram [Fig.~\ref{schem}(b)] for $\text{Pe}\ll 1$. We first summarize the equilibrium phase diagram [Fig.~\ref{narrow}(a), $\text{Pe}=0$] which lies within the universality class of Model C~\cite{agranov_thermodynamically_2024, bray_theory_1993} and describes ferromagnetic liquids~\cite{blume_ising_1971, Dauchot2019, Dauchot2020}. The stable phases are found by minimizing the free energy [Eq.~\eqref{free}] under the constraint of conserved density $(1/\ell_s)\int_0^{\ell_s} \rho dx = \rho_0$. At high temperatures ($\beta^{-1}>\beta_{\text{tri}}^{-1}$), a continuous transition line, separating the disordered homogeneous (H; $m=0$) and ordered homogeneous (H; $m\neq0$) states, terminates at a tricritical point. At low temperatures ($\beta^{-1}<\beta_{\text{tri}}^{-1}$), the transition between disordered and ordered states becomes discontinuous, and a \textit{static} phase separated state (P.S.) emerges with a coexistence between the two discrete symmetry phases ($m=0$ and $m\neq0$). The binodal curves, delineating the boundaries of the miscibility gap, follow from common-tangent construction and meet at the tricritical point~\cite{agranov_thermodynamically_2024}.

Any finite self-propulsion propulsion ($\Pe\neq0$) has a non-perturbative effect on the phase diagram topology [see for example Fig.~\ref{narrow}(b) which shows $\text{Pe}=0.08$].  In particular, the behaviour for high temperatures ($\beta^{-1}>\beta^{-1}_{\text{tri}}$) changes qualitatively, in that the equilibrium system ($\Pe=0$) features a critical line, but for $\Pe\neq0$ one sees two binodals separated by a miscibility gap. That is, the binodals that previously merged at $\beta^{-1}_{\text{tri}}$ now continue into the high-temperature regime.

Fixing $\Pe$, we denote the binodals $\rho_l(\beta)$ and $\rho_g(\beta)$ for the liquid and vapour, respectively; the corresponding spinodals are $\varphi_l(\beta)$ and $\varphi_g(\beta)$. We argue in \ref{ap.lowpe} that, for small Peclet, the gap between the spinodals scales as $\varphi_l(\beta)-\varphi_g(\beta)\sim \Pe$. We expect on general grounds that the gap between the binodals has the same scaling, which leads (see again \ref{ap.lowpe}) to\footnote{The notation $A\sim \Pe^\alpha$ has the same meaning as $A={\cal O}(\Pe^\alpha)$.}
\begin{equation}\label{expand0}
	(\rho_l - \rho_g) \sim \text{Pe},
	\quad
	m_l \sim \text{Pe}^{1/2} ,
	\quad V \sim \text{Pe}^{1/2} ,
	\qquad
	\text{for} \quad \beta^{-1} > \beta^{-1}_{\rm tri} ,
\end{equation}
where $m_l$ is the magnetization of the liquid phase in the travelling band. Note that the band velocity $V\sim \Pe^{1/2}$ is much faster than the typical speed of an individual particle (which scales as $\Pe$): this collective enhancement of velocity is possible because of the small density difference between the phases. Also, $m_l \gg (\rho_l-\rho_g)$ implies that the density of the minority species in the liquid phase is reduced with respect to the gas. Fig.~\ref{narrow}(c) shows that these scalings agree with numerical simulations of the deterministic hydrodynamics.

\begin{figure}
	\centering
	\includegraphics[width=.34\columnwidth]{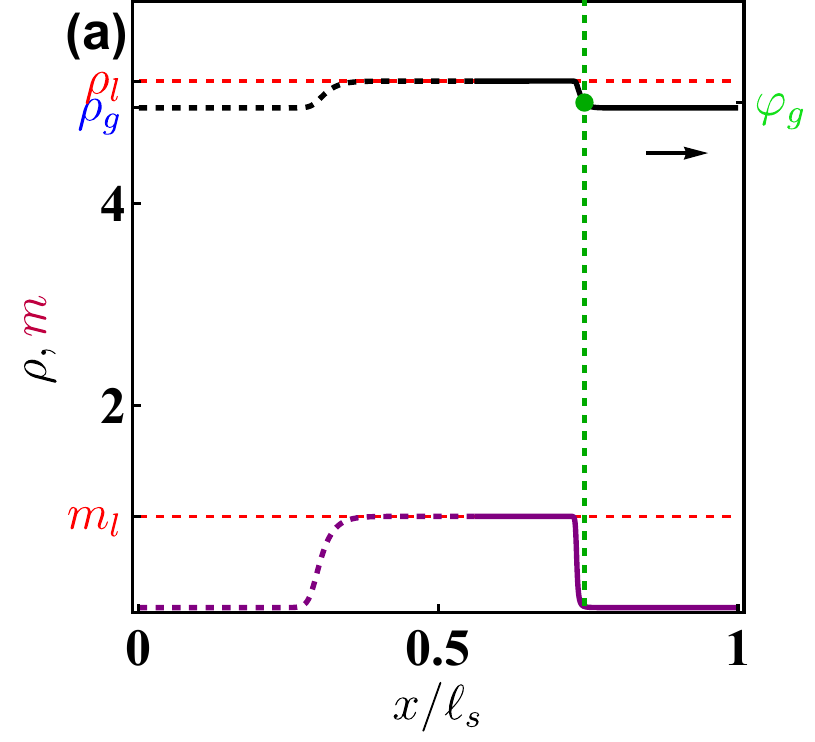}
		\includegraphics[width=.31\columnwidth]{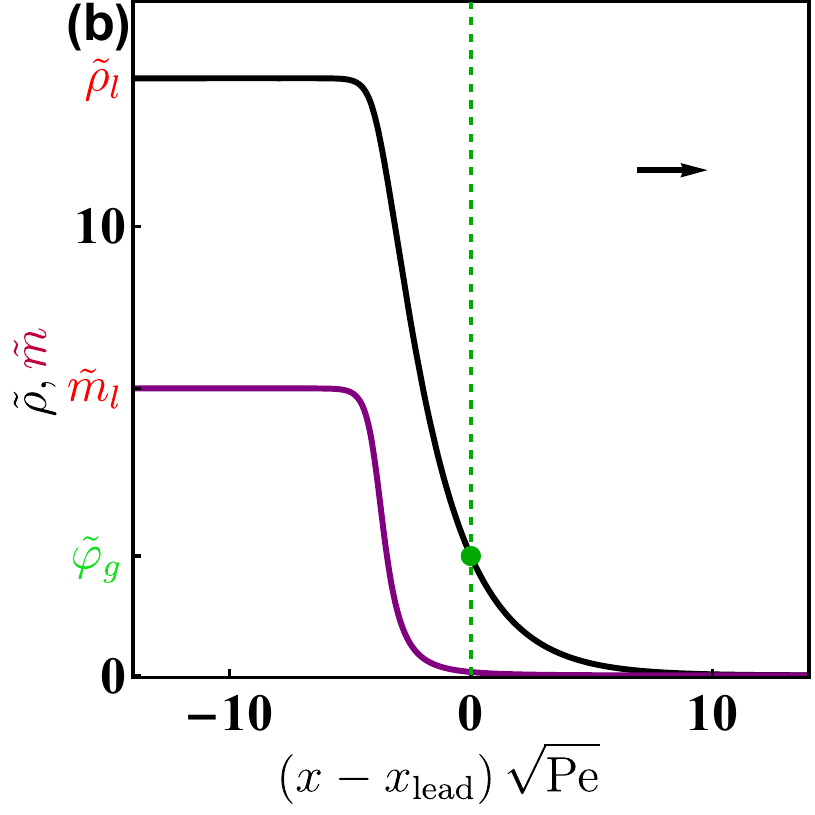}
	\includegraphics[width=.31\columnwidth]{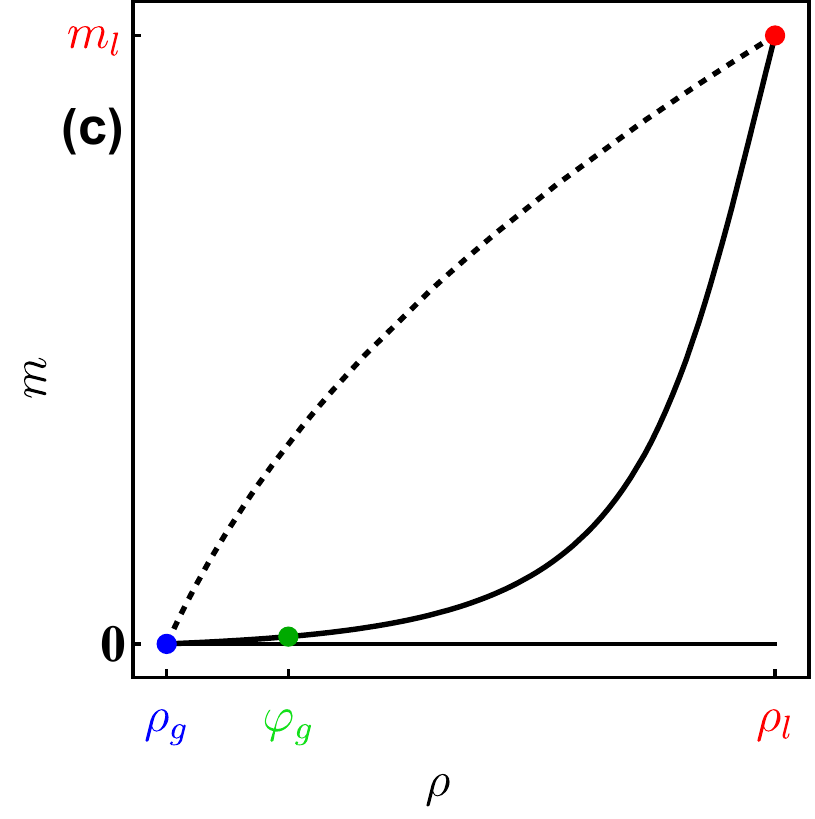}	
	\caption{(a)~Profiles of density and magnetization with the leading and trailing interfaces in solid and dashed lines, respectively. 
	 The green dot marks the point where the interfacial profile crosses the spinodal values $(\rho = \varphi_g, m\simeq 0)$.
	 (b)~Plot of the scaled profiles [Eq.~\eqref{equ:travel-scale-main}] displaying the expected scaling at low $\text{Pe}$. The value of $x_{\text{lead}}$ here corresponds to intersection with the gas spinodal. Parameters: $\text{Pe}=0.02$, $\rho_0\simeq5.1$, $\beta^{-1}=0.8$. (c)~Parametric representation of the same profile $\left[\rho(x,t),m(x,t)\right]$ with $x\in[0,\ell_s)$.}
	\label{lowpe}
\end{figure}

The travelling profiles [Eq.~\eqref{eq:travel}] feature some interfaces which diverge like $\Pe^{-1/2}$, see \ref{ap.lowpe}; consistently, the equilibrium model ($\Pe=0$) entails a second-order phase transition for $\beta^{-1}>\beta^{-1}_{\rm tri}$. For any fixed $\Pe>0$, one can still consider system sizes large enough that the travelling band is dominated by bulk liquid and vapor phases. The two interfaces have different behaviours, referred to as the leading and trailing edges, since the T.B. breaks left-right symmetry. For instance, Fig.~\ref{lowpe} shows a band with $V>0$ so the leading edge is the rightmost interface, which generically behaves as
\begin{equation}
	\begin{aligned}
	\label{equ:travel-scale-main}
	\bar\rho(x) &= \rho_g + \Pe \, \tilde\rho_{\rm lead}\!\left( (x-x_{\rm lead})\sqrt{\Pe}\right) + {\cal O}(\Pe^2) ,	
	\\
	\bar m(x) &= \sqrt{\Pe} \tilde m_{\rm lead}\!\left( (x-x_{\rm lead})\sqrt{\Pe}\right) + {\cal O}(\Pe^{3/2}) ,
	\end{aligned}
\end{equation}
where $x_{\rm lead}$ is the position of the leading interface (set arbitrarily), and $(\tilde\rho_{\rm lead},\tilde m_{\rm lead})$ are scaling functions of order unity. Profiles with $V<0$ are obtained by reflecting $(\bar\rho,\bar m$), and a result analogous to Eq.~\eqref{equ:travel-scale-main} holds for the trailing edge.

Remarkably, we can analyse the details of the functions $(\tilde\rho_{\rm lead},\tilde m_{\rm lead})$ via a mapping to a dynamical system (see~\ref{ap.lowpe}), where spatial profiles are described as heteroclinic trajectories between two fixed points~\cite{caussin_emergent_2016}. In particular, $\tilde m_{\rm lead}$ features a singularity at the location $x=x_{\rm sing}$ where the density reaches the value of the gas spinodal ($\tilde \rho_{\rm lead}=\varphi_g$) [see green dot in Fig.~\ref{lowpe}], so that $\tilde m_{\rm lead}=0$ for $x>x_{\rm sing}$, and $\tilde m_{\rm lead}>0$ for $x<x_{\rm sing}$. Therefore, at the front of the leading edge ($x>x_{\rm sing}$), there exists a region where, although the density already differs from the gas $[\rho(x) - \rho_g] \sim {\rm Pe}$, the magnetisation $m(x>x_{\rm sing}) \sim \Pe^{3/2}$ is much smaller than at the back of the leading edge $m(x<x_{\rm sing})\sim \Pe^{1/2}$. By contrast, at the trailing edge, the magnetisation is positive everywhere, thus smoothly connecting the interface to the bulk phase.


\section{Entropy production rate from particles to fields}\label{secentro}

We now turn to the EPR of the flocking model. The microsocopic definition of the model obeys local detailed balance [Sec.~\ref{lattice}]. A given trajectory $\omega$ is prescribed by the time series of particles' hops and their spin flips. The (time-averaged) rate of entropy production $S_{\rm micro}$ for a trajectory $\omega$ of duration $T$ can then be identified in terms the log-probability ratio of trajectory $\omega$ and its time-reversed counterpart $\omega^{R}$~\cite{seifert_stochastic_2012, maes_local_2021}:
\begin{equation}\label{epr}
	S_{\rm micro}(\omega) = \frac{1}{T} \ln\frac{P\left(\omega\right)}{P\left(\omega^R\right)} .
\end{equation}
From local detailed balance, this EPR can be identified with a dissipated heat along the trajectory~\cite{seifert_stochastic_2012}.

On the other hand, a trajectory ${\cal X}$ at the hydrodynamic level [Eq.~\eqref{pathX}] is given by realizations of the relevant fields and currents. Its stochastic entropy production is defined by comparing $\cal X$ with its time-reversed counterpart ${\cal X}^{\rm R}$ as 
\begin{equation}\label{iepr}
	S_{\rm hydro}({\cal X}) = \frac{1}{T} \ln \frac{P({\cal X})}{P({\cal X}^R)} .
\end{equation}
Since information about individual particle trajectories is lost on taking the hydrodynamic limit, there is no guarantee that $S_{\rm hydro}$ should coincide with $S_{\rm micro}$ a priori~\cite{fodor_irreversibility_2022,obyrne_time_2022}. The former is sometimes referred to as the {\em informatic} entropy production rate (IEPR) to reflect that it is defined directly in terms of trajectory probabilities, and is not generically related to dissipated heat or thermodynamic entropy. This IEPR is generally smaller than its microscopic counterpart~\cite{pruessner_field_2022, cocconi_scaling_2022}. However, we show below that the hydrodynamic IEPR and the microscopic EPR actually coincide for the our flocking model.


\subsection{Quantifying EPR from microscopic lattice dynamics}\label{sec:epr_micro}

From the microscopic definition of the particle model [Eq.~\eqref{equ:wi}], the entropy production~\cite{seifert_stochastic_2012} associated with a hop to the right by a particle of type $\sigma$ is 
\begin{equation}
\Delta S_{\rm hop} = \ln \frac{ w_\sigma(i\to i+1) }{ w_\sigma(i+1\to i) } = \frac{\sigma\lambda}{LD_0} - \beta \Delta H \; .
\label{equ:dS}
\end{equation}
Similarly, for particles changing their orientation, the entropy production is $\Delta S_{\rm flip} =-\beta \Delta H$. In the absence of self-propulsion ($\lambda=0$), the total entropy production simply adds up to the energy difference:
\begin{equation}\label{eqent}
	S_{\rm micro}^{(\lambda=0)}(\omega)	= \frac{\beta}{T} \left[ H(\omega_0) - H(\omega_T) \right] ,
\end{equation} 
where $(\omega_0,\omega_T)$ are the configurations at the start and end points of trajectory $\omega$. For systems initialised in their stationary state, the average of $H$ is independent of time, so that $S_{\rm micro}^{(\lambda=0)}$ is zero on average. For general initial conditions, we still get $S_{\rm micro}^{(\lambda=0)}\to0$ as $T\to\infty$, provided that $H$ is bounded.

With self-propulsion ($\lambda\neq0$), the additional contribution in Eq.~\eqref{equ:dS} leads to
\begin{equation}\label{micro1}
	S_{\rm micro}(\omega) = \frac{\lambda}{D_0 L T} \left[ N_+(\omega) - N_-(\omega) \right] +	\frac{\beta}{T} \left[ H(\omega_0) - H(\omega_T) \right] ,
\end{equation} 
where $N_\sigma(\omega)=N^{\rm R}_\sigma(\omega) - N^{\rm L}_\sigma(\omega)$, in which $N^{\rm R}_\sigma(\omega)$ is the total number of right-hops by particles of type $\sigma$, and similarly $N^{\rm L}_\sigma(\omega)$ is the number of left-hops. That is, $N_\sigma(\omega)$ is the net displacement due to hops  by particles of type $\sigma$. These displacements can be expressed in terms of the particle currents as
\begin{equation}\label{fn}
	N_{\sigma}(\omega) = \frac{L^{2}}{\ell_s^2}\int_0^{\ell_s}dx\int_0^T\!dt\, J_{\sigma}(\omega;x,t) ,
\end{equation}
where $J_{\sigma}(\omega;x,t)$ is the current for particles of type $\sigma$, evaluated for the trajectory $\omega$ of the microscopic model. ( Eq.~\eqref{fn} provides a link between microscopic and hydrodynamic representations, the {$L/\ell_s$}-dependence reflects that $\rho$ is the number of particles per site, recall Sec.~\ref{hyd}.) Hence, using the definitions of the Péclet number [Eq.~\eqref{equ:xi-xi-pe}] and the flux $J_m = J_+ - J_-$ [Eq.~\eqref{d1}], we obtain
\begin{equation}\label{micro}
	S_{\rm micro}(\omega) =	\frac{L \text{Pe}}{\ell_sT}\int_0^{\ell_s}\!dx\int_0^T\!dt \,J_m(\omega;x,t) + \frac{\beta}{T} \left[ H(\omega_0) - H(\omega_T) \right]  .
\end{equation}
We emphasize that this expression gives the rate at which entropy is produced at the particle level~\cite{seifert_stochastic_2012, pietzonka_entropy_2018}. Multiplying Eq.~\eqref{micro} by the temperature $\beta^{-1}$ yields
\begin{equation}\label{eq:1st}
	Q = W + \Delta H ,
\end{equation}
where $Q=T\beta^{-1} S_{\rm micro}$ is the dissipated heat, $W = \Pe (L/\ell_s) \int dx dt J_m $ is the work done by the self-propulsion, and $\Delta H$ is the change in energy. The relation in Eq.~\eqref{eq:1st} is analogous to the first law of thermodynamics: it reflects the conservation of energy between the system (namely, active particles), the surrounding thermostat which absorbs heat, and some underlying reservoirs (for instance, chemostats) which provide work to power self-propulsion.


\subsection{Quantifying EPR from fluctuations of hydrodynamic fields}
\label{sec:epr-hydro}

To evaluate the hydrodynamic EPR [Eq.~\eqref{iepr}], one has to specify the parity under time reversal of the observable fields when defining the reversed trajectory ${\cal X}^R$~\cite{fodor_irreversibility_2022}. For consistency with the microscopic EPR, we take particles identities to be invariant under time reversal. Accordingly, density ($\rho$) and magnetization ($m$) are even, and Eq.~\eqref{d1} enforces that the fluxes $(J_\rho,J_m,K)$ are odd:
\begin{equation}\label{negexttr}
	{\cal X}^R = \left\{\rho(x,T-t),m(x,T-t),-J_{\rho,m}(x,T-t),-K(x,T-t) \right\}_{x\in[0,\ell_s],t\in[0,T]} .
\end{equation} 
Note that a different time reversal can be chosen for other dynamics~\cite{fodor_irreversibility_2022}. For instance, in flocking systems where the positions of particles are enslaved to their orientations~\cite{ferretti_signatures_2022}, namely without any noise terms, then $m$ is odd and, correspondingly, $(J_m,K)$ are even. In fact, in previous studies of EPR in field theories of flocking, particle orientations and hydrodynamic magnetization were treated as odd~\cite{borthne_time-reversal_2020-1}.

From the expression of the path action ${\cal A}$ [Eq.~\eqref{equ:path-LDP}], substituting the choice for ${\cal X}^R$ [Eq.~\eqref{negexttr}] into the definition of the hydrodynamic EPR [Eq.~\eqref{iepr}] yields 
\footnote{The proper evaluation of entropy production using a field theory requires a careful definition of time discretization, say It\=o or Stratonovich~\cite{cates_stochastic_2022}. However, the scaling of fluctuation magnitude with $L^{-1/2}$ implies that this choice is not important here.}
\begin{equation}\label{entro0}
	 S_{\rm hydro}({\cal X}) = \frac{L}{\ell_sT}\int_0^T dt\int_0^{\ell_s} dx
	\left\{2\begin{bmatrix}
	J_{\rho}\\
	J_m
	\end{bmatrix}^\dag
	\mathbb{C}^{-1}	
	\begin{bmatrix}
	\bar{J}_{\rho}\\
	\bar{J}_m 
	\end{bmatrix}+ 2K\frac{\partial\mathcal{F}}{\partial m}\right\} ,
\end{equation}
where we have used the explicit expressions of the Lagrangians [Eqs.~\eqref{fluxrate} and ~\eqref{phi12}] and the mean fluxes [Eq.~\eqref{d2}]. It is convenient to define the hydrodynamic flux $j$ and force $f$ as~\cite{kaiser_canonical_2018}
\begin{equation}\label{jf}
	j = \begin{bmatrix}
	J_{\rho}\\
	J_{m}\\
	2K
	\end{bmatrix}\quad,\quad f=\begin{bmatrix} \bar{f}_\rho 
	\\ \bar{f}_m
	\\
	\frac{\partial\mathcal{F}}{\partial m}
	\end{bmatrix} \quad, \qquad \text{with} \quad
	\begin{bmatrix} \bar{f}_\rho \\ \bar{f}_m \end{bmatrix} = 2\mathbb C^{-1} \begin{bmatrix}\bar{J}_{\rho}\\\bar{J}_{m}\end{bmatrix}
\end{equation}
Then the EPR in Eq.~\eqref{entro0} becomes
\begin{equation}\label{entro3}
	S_{\rm hydro}({\cal X}) = \frac{L}{\ell_sT}\int_0^T \!dt\int_0^{\ell_s} \!dx\,
	{(j\cdot f)} .
\end{equation}
We now derive a hydrodynamic {counterpart} of Eq.~\eqref{micro}. It is useful to separate the force $f$ into into equilibrium and active contributions, with
\begin{equation}\label{feqac}
	f = f_{\text{eq}}+f_{\text{active}},
	\quad
	f_{\text{eq}} = 
	\begin{bmatrix}
		-\partial_x\frac{\partial\mathcal{F}}{\partial\rho}\\
		-\partial_x\frac{\partial\mathcal{F}}{\partial m}\\
		\frac{\partial\mathcal{F}}{\partial m}
	\end{bmatrix} ,
	\quad
	f_{\text{active}} = 
	\begin{bmatrix}
		0\\
		\text{Pe}\\
		0
	\end{bmatrix} .
\end{equation}
First, considering the contribution of $f_{\rm eq}$ in the hydrodynamic EPR [Eq.~\eqref{entro3}], we get
\begin{equation}\label{eq}
\begin{aligned}
	\frac{L}{\ell_sT}\int_0^T dt\int_0^{\ell_s} \! dx \, (j\cdot f_{\text{eq}})   &= - \frac{L}{\ell_sT}\int_0^T dt\int_0^{\ell_s} dx \bigg( \frac{\partial\mathcal{F}}{\partial\rho} \partial_t \rho + \frac{\partial\mathcal{F}}{\partial m} \partial_t m \bigg)
	\\
	&= \frac{F_0- F_T}{T} ,
\end{aligned}
\end{equation}
where $F_T$ is the free energy [Eq.~\eqref{eq:FF}] evaluated at time $T$, and similarly $F_0$ is evaluated at time $0$. The first equality in Eq.~\eqref{eq} uses an integration by parts as well as the equation of motion Eq.~\eqref{d1}; the second is the chain rule. Physically, Eq.~\eqref{eq} illustrates that equilibrium forces generate entropy equal to the free-energy difference. The active contribution to Eq.~\eqref{entro3} is
\begin{equation}\label{entro4}
	\frac{L}{\ell_sT}\int_0^T dt\int_0^{\ell_s} \!dx\, (j\cdot f_{\text{active}})  = \frac{L \text{Pe}}{\ell_sT}\int_0^T dt\int_0^{\ell_s} \!dx \,J_m .
\end{equation}
Hence, Eq.~\eqref{entro3} becomes
\begin{equation}
	S_{\rm hydro}({\cal X}) = \frac{L \text{Pe}}{\ell_sT}\int_0^T dt\int_0^{\ell_s} \!dx \,J_m + \frac1T [F_0- F_T] .
\label{entro-X}
\end{equation}
Comparing Eq.~\eqref{entro-X} with Eq.~\eqref{micro}, the first term matches exactly (it corresponds to rate of work done by self-propulsion); the second term differs in that the energy $H$ is replaced by the free energy $F$. Recall that $F=\beta H - S_{\rm sys}$ is actually dimensionless, where $S_{\rm sys}$ refers to the system entropy which drives diffusive spreading of the density.

In the following, we focus on the steady-state average $\langle\cdot\rangle$ of EPR, for which the terms involving $(H,F)$ in Eqs.~(\ref{micro},\ref{entro-X}) vanish. The relevant time-integrals are also stationary in this regime, yielding
\begin{equation}
\label{equ:ave-ep}
	\langle S_{\rm hydro} \rangle = \langle S_{\rm micro} \rangle = \frac{L \text{Pe}}{\ell_s}\int_0^{\ell_s} \!dx \,\bar{J}_m ,
\end{equation}
with $\bar{J}_m$ as in Eq.~\eqref{d2}, evaluated in steady state. In particular, $\bar{J}_m$ is a function of $x-Vt$ for T.B. states [Eq.~\eqref{eq:travel}], so that $\int dx \bar{J}_m$ is indeed independent of time.  It is unusual that the microscopic EPR gets preserved under coarse-graining of the dynamics~\cite{fodor_irreversibility_2022}. A similar result has been established for a related lattice gas model of active phase separation~\cite{agranov_entropy_2022}. In our flocking model, the correspondence between microscopic and hydrodynamic EPR essentially relies on (i)~the proper scaling of the microscopic jump rates with system size [Fig.~\ref{schem}], which allows for a controlled coarse-graining, and (ii)~a thermodynamically consistent definition of such rates, such that the same microscopic energy $H$ constrains both particle hops and spin flips. In fact, the outcome of (i) is that typical particle trajectories span the hydrodynamic system size on hydrodynamic time scales, so that the coarse-grained hydrodynamic equations become exact at large system sizes: it yields a correspondence between the path actions of the hydrodynamics and the underlying microscopic process~\cite{kaiser_canonical_2018}.



\subsection{Spatial decomposition of EPR: Where dissipation matters}\label{sec:ave-epr}

The spatial integrand in the microscopic EPR [Eq.~\eqref{micro}], which stems from summing over the microscopic hops [Eq.~\eqref{equ:dS}], provides a natural decomposition of EPR. Using the correspondence between microscopic and hydrodynamic EPR, we examine the spatial decomposition of EPR using the hydrodynamic representation [Eq.~\eqref{entro3}]. In fact, such a decomposition of the hydrodynamic EPR allows one to delineate where the contributions of the microscopic hops to EPR occur in space.

In what follows, we focus on the steady-state averaged EPR $\langle S_{\rm hydro}\rangle$, for which $\jbar\cdot f$ is the relevant local decomposition. Here, the mean current $\jbar$ [Eq.~\eqref{jf}] is obtained by replacing $(J_\rho,J_m,K)$ with their most probable values $(\bar J_\rho,\bar J_m,\bar K)$ [Eq.~\eqref{d2}], which explicitly depend on $(\rho,m)$.

\subsubsection{EPR across phase diagram: Maximal dissipation in homogeneous states.}

The local decomposition of EPR $\jbar\cdot f$ provides a route to anticipate how the {spatially} integrated EPR $\langle S_{\rm hydro} \rangle$ behaves in various states of the system. To this end, we further decompose the current as 
\begin{equation}
	\jbar = \jbar_{\rm eq} + \jbar_{\rm active} ,
\end{equation}
with 
\begin{equation}
\label{jbareqact}
 \jbar_{\text{eq}}=
  \begin{bmatrix}
   \jbar_{{\rm eq},\rho} \\ \jbar_{{\rm eq},m}
  \\
  2M\sinh\left(\frac{\partial\mathcal{F}}{\partial m}\right)
  \end{bmatrix} ,
  \quad\quad
  \jbar_{\text{active}}=\begin{bmatrix}
  \text{Pe} \, m\\
  \text{Pe}\,\rho\\
  0
  \end{bmatrix} \; ,
  \quad\quad  \text{where}\quad
  \begin{bmatrix} \jbar_{{\rm eq},\rho} \\ \jbar_{{\rm eq},m}\end{bmatrix} 
  = -\frac12 \mathbb C\begin{bmatrix}\partial_x\frac{\partial\mathcal{F}}{\partial\rho}\\\partial_x\frac{\partial\mathcal{F}}{\partial m}\end{bmatrix}
  \; .
\end{equation}
The currents $(\jbar_{\rm eq},\jbar_{\rm active})$ are canonically conjugate~\cite{maes_cano_2008,kaiser_canonical_2018} to the forces $(f_{\rm eq},f_{\rm active})$ introduced in Eq.~\eqref{feqac}. The local EPR then decomposes into three contributions
\begin{equation}
\jbar \cdot f = s_{\rm bulk} + s_{\rm interface} + s_{\rm rev}
\end{equation}
where
\begin{equation}\label{decomp}
	s_{\text{bulk}}= \jbar_{\text{active}}\cdot f_{\text{active}} \; ,
	\quad\quad
	s_{\text{interface}} = \jbar_{\text{eq}}\cdot f_{\text{active}} \;,
	\quad\quad
	s_{\rm rev} = \jbar \cdot f_{\rm eq} \; .
\end{equation}
Using Eqs.~(\ref{feqac},\ref{jbareqact}), we deduce
\begin{equation}\label{bulk}
	\begin{aligned}
	\frac{L}{\ell_s }\int_0^{\ell_s} \!dx\, s_{\text{rev}} &= \frac{1}{T} \big[ F_0 - F_T \big] ,
	\\
	\frac{L}{\ell_s }\int_0^{\ell_s} \!dx\, s_{\text{bulk}} &=	\frac{L\text{Pe}^2}{\ell_s} \int_0^{\ell_s} dx \rho = N\text{Pe}^2 ,
	\end{aligned}
\end{equation}
where $N=L\rho_0$ is the total number of particles, and we have used mass conservation $\int_0^{\ell_s}dx\rho/\ell_s=\rho_0$. Therefore, the integrated  contribution of $s_{\rm rev}$ leads to the free-energy difference [Eq.~\eqref{eq}]. Although the space integral of $s_{\rm rev}$ vanishes in steady state, it can be locally non-zero; for instance, $s_{\rm rev}>0$ indicates that the local current $\jbar$ flows down the free energy gradient (parallel to $f_{\rm eq}$). In contrast, the integrated contribution of $s_{\rm bulk}= \Pe^2\rho$ is non-zero. Moreover, the interfacial contribution $s_{\text{interface}}$ vanishes in homogeneous regions, since the first two components of the equilibrium flux $\bar \jmath_{\rm eq}$ [Eq.~\eqref{jbareqact}] are proportional to gradients of $(\rho,m)$.

The bulk and interfacial contributions add up to give the local rate of work:
\begin{equation}
	s_{\rm bulk} + s_{\rm interface} = \Pe\, \bar J_m  ,
\end{equation}
and yield the steady-state EPR when integrated over space:
\begin{equation}\label{epr_decomp}
	\langle S_{\rm hydro} \rangle =  \frac{L}{\ell_s} \int_0^{\ell_s} dx\left( s_{\text{interface}} + s_{\text{bulk}}\right) .
\end{equation}
It follows that, for homogeneous states, the EPR is given by $\langle S_{\rm hydro} \rangle = N \Pe^2$. We now demonstrate that $N \Pe^2$ is an upper bound on the EPR for all steady states. First, observe from Eqs.~(\ref{feqac},\ref{jbareqact}) that
\begin{equation}
\label{jf-recip}
	\jbar_{\text{eq}}\cdot f_{\text{active}} = \jbar_{\text{active}}\cdot f_{\text{eq}} ,
\end{equation}
yielding
\begin{equation}
	s_{\text{interface}} = \jbar_{\text{active}}\cdot f_{\text{eq}} = s_{\rm rev} - \jbar_{\text{eq}}\cdot f_{\text{eq}} .
\label{equ:sint}
\end{equation}
Second, we deduce from Eqs.~(\ref{feqac},\ref{jbareqact}) that
\begin{equation}
	 \jbar_{\text{eq}}\cdot f_{\text{eq}} = \frac{1}{2}  \begin{bmatrix}
	\partial_x\frac{\partial \mathcal{F}}{\partial\rho}\\
	\partial_x\frac{\partial \mathcal{F}}{\partial m}
	\end{bmatrix}^\dag
	\mathbb C\begin{bmatrix}
	\partial_x\frac{\partial  \mathcal{F}}{\partial\rho}\\
	\partial_x\frac{\partial  \mathcal{F}}{\partial m}
	\end{bmatrix}+2M\frac{\partial\mathcal{F}}{\partial m}\sinh\left(\frac{\partial\mathcal{F}}{\partial m}\right)
	\geq 0 ,
	\label{eq:in}
\end{equation}
where we have used that $(M,\mathbb{C})$ are non-negative. Combining Eqs.~(\ref{equ:sint},\ref{eq:in}), and given that the integrated contribution of $s_{\rm rev}$ vanishes, it follows that the integrated contribution of $s_{\rm interface}$ is always negative:
\begin{equation}
\frac{L}{\ell_s} \int\!dx\, s_{\rm interface} = \int\!dx\, (-\jbar_{\rm eq} \cdot f_{\rm eq}) \leq 0 \; .
\label{equ:sint-neg}
\end{equation}
Note however that $s_{\rm interface} \neq -\jbar_{\rm eq} \cdot f_{\rm eq}$ in general: these two quantities differ locally but have the same integral. Finally, using Eq.~\eqref{equ:sint-neg} with Eqs.~(\ref{epr_decomp},\ref{bulk}), we obtain
\begin{equation}\label{ent5}
	0 \leq \langle S_{\rm hydro} \rangle \leq N {\rm Pe}^2 ,
\end{equation}
where the first equality is the second law of thermodynamics. In homogeneous states, the upper bound becomes an equality; the lower bound is an equality in equilibrium.

In short, the system dissipates most in homogeneous states, and the emergence of T.B. reduces the dissipation rate. Remarkably, the value of EPR in the homogeneous states coincides with the case of non-interacting systems\footnote{A typical particle produces $\Pe^2 k_B$ of entropy during its orientational relaxation time. To see this, use Eq.~\eqref{equ:t-hydro} and that the entropy production is measured in units of $k_{\rm B}$.}: it is independent of whether the system is in the disordered state ($m=0$) or in collective motion ($m\neq0$). This result is in stark contrast with other flocking models which entail a peak of EPR at the transition between disorder and C.M.~\cite{yu_energy_2022, ferretti_signatures_2022, ferretti_out_2024}. In fact, such a peak is present in these models even in the absence of self-propulsion: it is not related to particles' self-propulsion, but arises simply because the diffusive and orientational dynamics are not thermodynamically consistent with each other. Even in the absence of self-propulsion, these models do not obey detailed balance with respect to any Hamiltonian $H$, contrary to our flocking model.


\subsubsection{Modulation of EPR at leading and trailing edges.}\label{sec:interf-epr-local}

From Eqs.~(\ref{feqac},\ref{jbareqact}), we write the reversible EPR as
\begin{equation}
s_{\rm rev} = [ 2\bar K + \partial_x \bar J_m ] \frac{\partial {\cal F}}{\partial m} +  \partial_x \bar J_m  \frac{\partial {\cal F}}{\partial \rho}  - \partial_x \left[  \bar J_\rho   \frac{\partial {\cal F}}{\partial \rho} + \bar J_m   \frac{\partial {\cal F}}{\partial m} \right] .
\end{equation}
In the T.B. state, Eq.~\eqref{d1} holds with $(K,J_\rho,J_m)$ equal to their barred values, so the chain rule gives
\begin{equation}
	s_{\rm rev} = -\frac{\partial {\cal F} }{\partial t} - \partial_x \left[ \bar J_\rho \frac{\partial {\cal F}}{\partial \rho} + \bar J_m  \frac{\partial {\cal F}}{\partial m} \right] .
\end{equation}
For a travelling profile [Eq.~\eqref{eq:travel}], we have $\partial {\cal F}/\partial t = -V \partial_x {\cal F}$, yielding
\begin{equation}\label{eq:rev}
	s_{\rm rev} =  \partial_x \left[ V{\cal F} - \bar J_\rho   \frac{\partial {\cal F}}{\partial \rho} 
- \bar J_m   \frac{\partial {\cal F}}{\partial m} \right] .
\end{equation}
Since Eq.~\eqref{eq:rev} is a total derivative, we deduce $\int_0^{\ell_s} \!dx\, s_{\rm rev}=0$ as expected. Locally, $s_{\rm rev}$ has contributions from leading and trailing edges of the travelling band.  The total contribution from the leading edge can be obtained by integrating across it to obtain
\begin{equation}
\begin{aligned}
	S_{\rm rev}^{\rm leading} & = \frac{L}{\ell_s} \int_{\rm liq}^{\rm gas} \!dx\, s_{\rm rev} 
	\\
	& = \frac{L}{\ell_s} \left[ V {\cal F}  - \bar J_\rho \frac{\partial {\cal F}}{\partial \rho}  \right]_{\rm liq}^{\rm gas} ,
	\label{eq:Srev-lead}
\end{aligned}
\end{equation}
where we used that $(\partial {\cal F}/\partial m)=0$ in the bulk of either phase.\footnote{In the bulk, one has $\partial_x J_m=0$, from which Eq.~\eqref{d1} implies $K=0$, hence $(\partial {\cal F}/\partial m)=0$.} We also continue to assume $V>0$, the opposite case is analogous. For bulk phases one also has $\bar J_\rho = m\Pe$, the gas has $m=0$, and using Eq.~\eqref{eq:VV} we obtain 
\begin{equation}
\label{equ:SP}
	S_{\rm rev}^{\rm leading} = {\frac{L}{\ell_s}}V [ \Phi_{\rm gas} - \Phi_{\rm liq} ],
	\quad
	\Phi(\rho,m) = {\cal F}(\rho,m) - \rho \left( \frac{\partial {\cal F}}{\partial \rho} \right)_{\rm liq} .
\end{equation}
Since $s_{\rm rev}$ integrates to zero, one has trivially that $S_{\rm rev}^{\rm trailing} = - S_{\rm rev}^{\rm leading}$.

\begin{figure}
	\centering
	\includegraphics[width=.4\columnwidth]{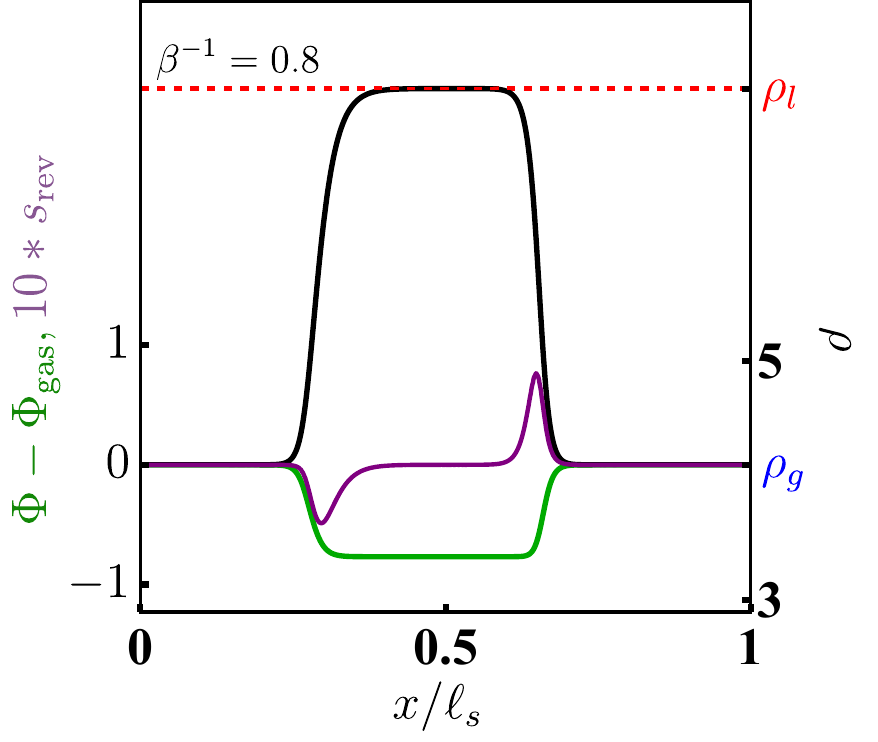}
	\caption{The reversible EPR density $s_{\rm rev}$ [purple line, Eq.~\eqref{eq:rev}] displays opposite peaks that are localized at interfaces of the density profile [black line]. The area beneath these localized modulations is given by the difference in $\Phi$ [green line, Eq.~\eqref{equ:SP}].
	Parameters: $\text{Pe}=1$, $\rho_0=5.25$.}
	\label{figphi}
\end{figure}

The behaviour of $s_{\rm rev}$ and $\Phi$ is shown in Fig.~\ref{figphi}.  This model has $\Phi_{\rm liq} < \Phi_{\rm gas}$, so $s_{\rm rev}$ is positive at the leading edge and negative at the trailing edge. Recalling the definition $s_{\rm rev} = \jbar \cdot f_{\rm eq}$ [Eq.~\eqref{decomp}], the leading edge ($s_{\rm rev}>0$) has the current $\jbar$ in the same direction as the thermodynamic force $f_{\rm eq}$, which points down the gradient of ${\cal F}$. At the trailing edge, one has the opposite effect: the current points up the gradient of ${\cal F}$, which increases (locally) the free energy and reduces dissipation.

The definition of $\Phi$ [Eq.~\eqref{equ:SP}] resembles (minus) the equilibrium thermodynamic pressure $-{\cal F} + \rho (\partial {\cal F}/\partial \rho)$. Note that replacing ${\cal F} \to {\cal F} - c\rho$ in the free-energy definition [Eq.~\eqref{free}] does not affect the dynamics, since $\int dx \rho$ is a conserved quantity. Hence, physical quantities cannot be ruled by local free-energy differences (e.g., ${\cal F}_{\rm liq} - {\cal F}_{\rm gas}$) which depend on $c$. The structure of \eqref{equ:SP} ensures that $\Phi$ is independent of $c$; {the similarity with the pressure reflects that both quantities obey this same constraint.}

\begin{figure}
	\centering
	\includegraphics[width=.31\columnwidth]{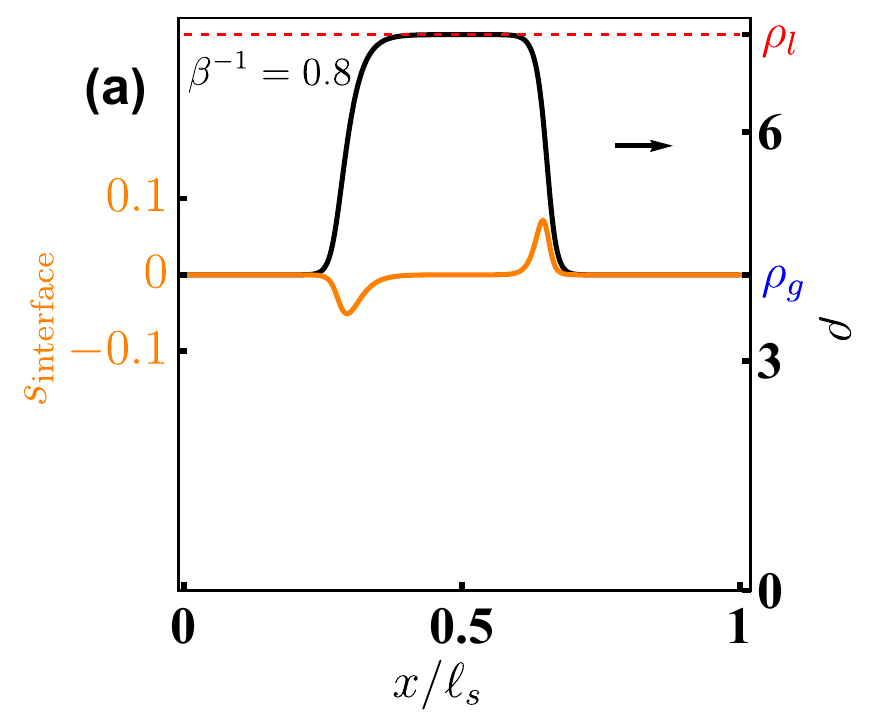}
	\includegraphics[width=.31\columnwidth]{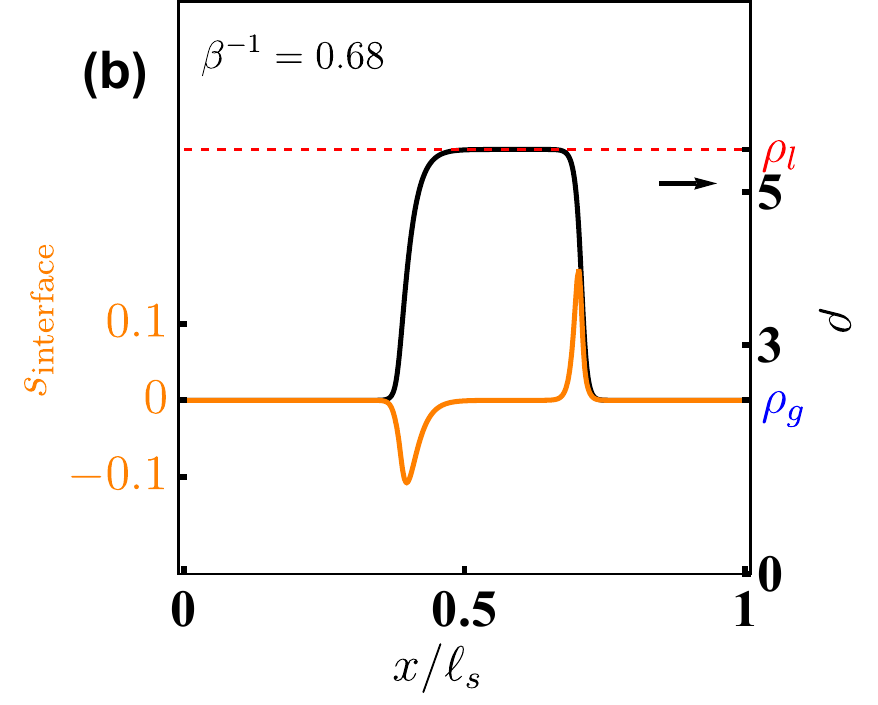}
	\includegraphics[width=.31\columnwidth]{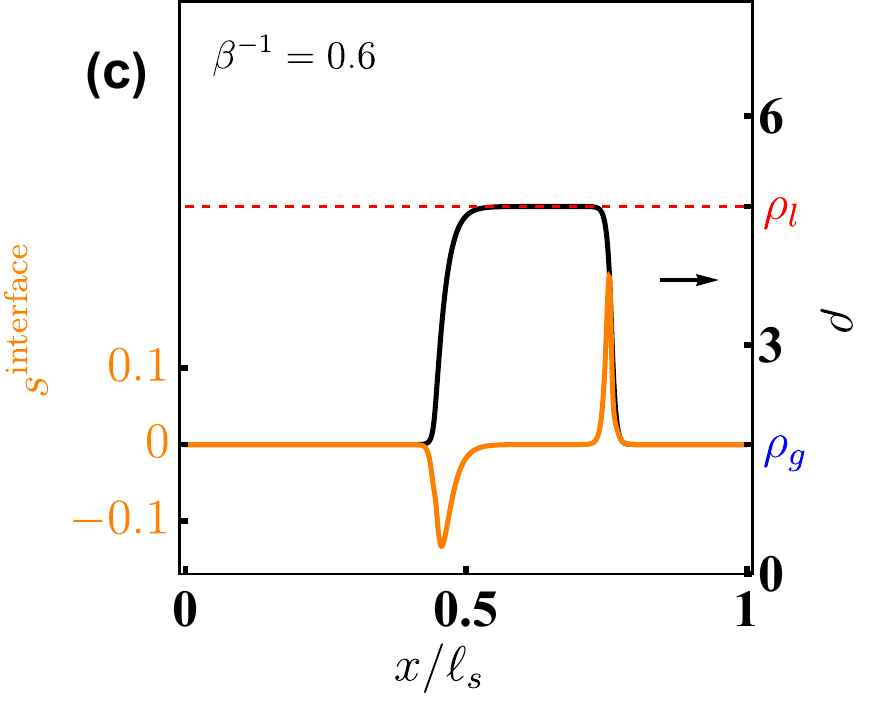}
	\\	
	\includegraphics[width=.31\columnwidth]{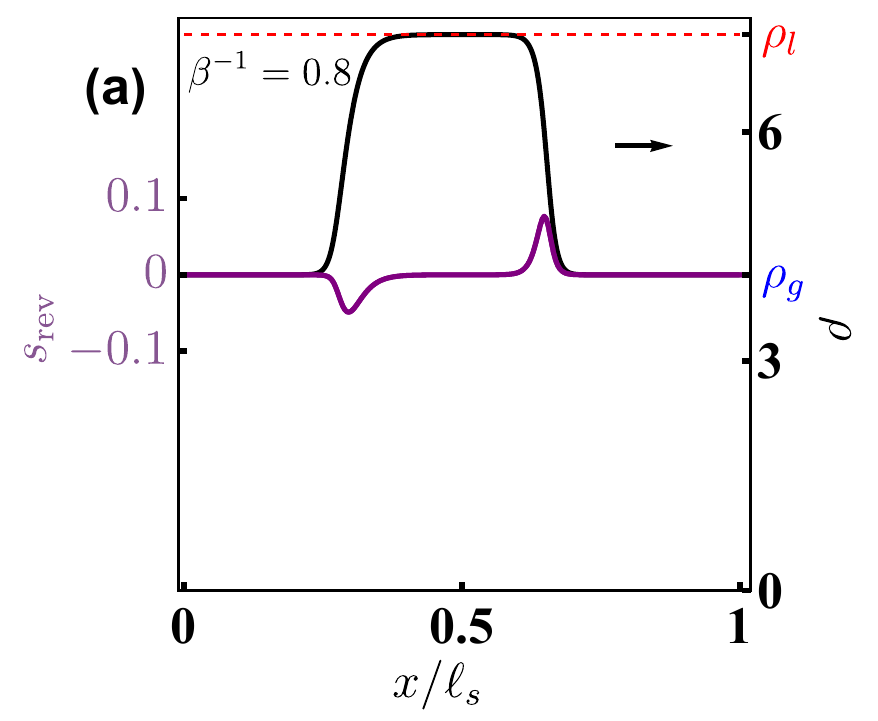}
	\includegraphics[width=.31\columnwidth]{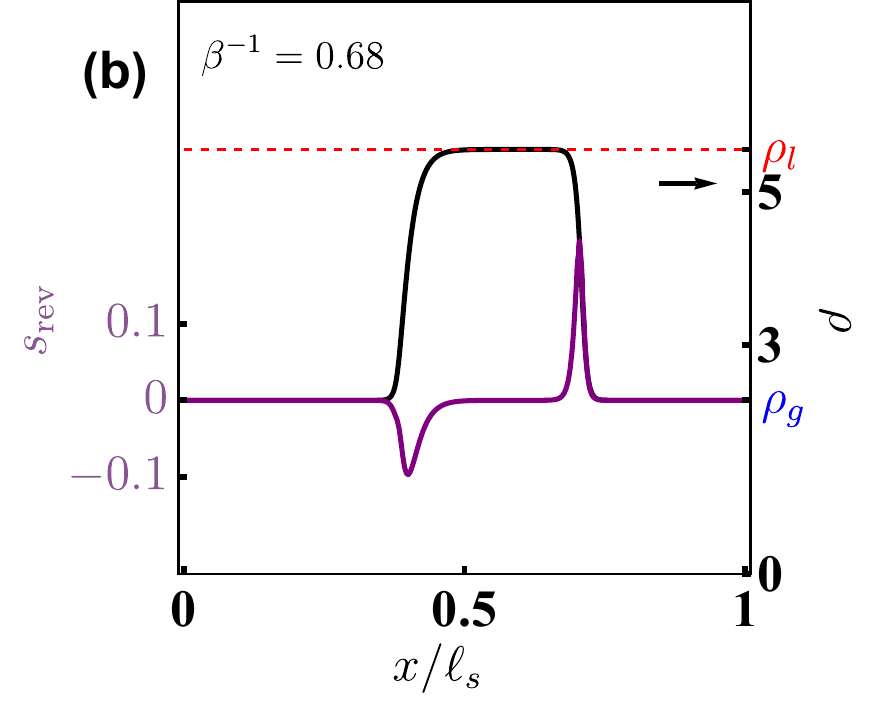}
	\includegraphics[width=.31\columnwidth]{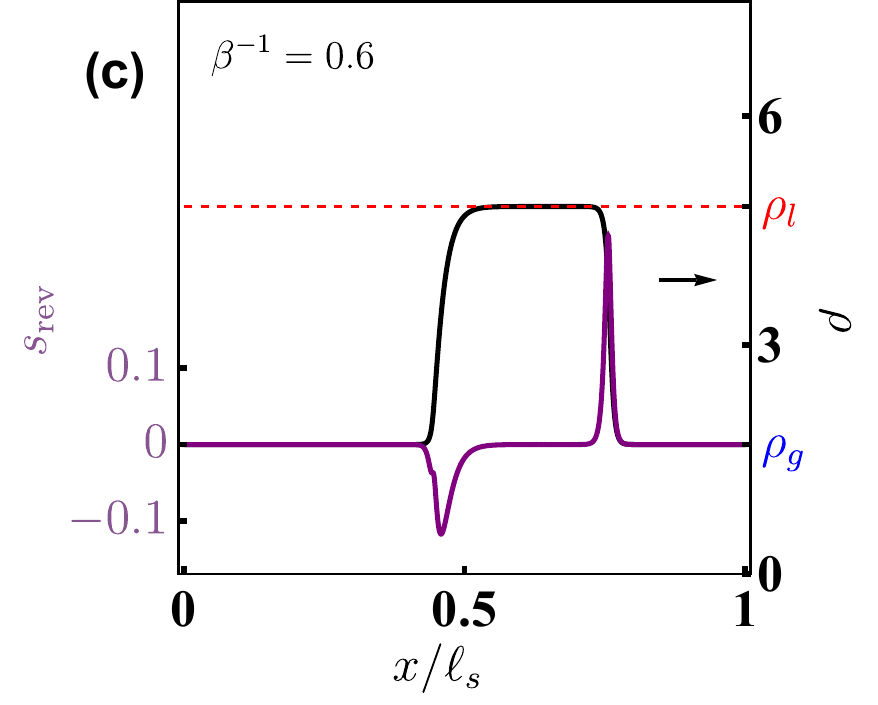}
	\caption{The interfacial EPR density $s_{\rm interface}$ [top row, orange line, Eq.~\eqref{equ:sint}] displays opposite peaks that are localized at interfaces of the density profile [black line]. It has a behavior similar qualitative to the reversible EPR density $s_{\rm rev}$ [bottom row, purple line, Eq.~\eqref{eq:rev}]. Here $\text{Pe}=1$ and the mean densities are $\rho_0=(5.25, 3.28, 2.62)$ in (a,b,c), respectively.}
	\label{interface}
\end{figure}

We now turn to the interfacial contribution $s_{\rm interface} = \jbar_{\rm eq}\cdot f_{\rm active} = \jbar_{\rm active}\cdot f_{\rm eq} $ [Eq.~\eqref{decomp}] to the local EPR. Figure~\ref{interface} shows that $s_{\rm interface}$ behaves similarly to $s_{\rm rev}$: it is positive at the leading edge and negative at the trailing edge. However, we recall that the contributions of the two interfaces do not cancel in this case, so that $\int dx s_{\rm interface}<0$ [Eq.~\eqref{equ:sint-neg}]. From Eqs.~(\ref{equ:sint}, \ref{eq:in}), we obtain $s_{\rm interface} \leq s_{\rm rev}$, and we also find numerically that  $s_{\text{bulk}}+s_{\text{interface}}\geq 0$ at every location in space. Note also that replacing ${\cal F}$ by $\Phi$ in Eq.~\eqref{feqac} leaves the thermodynamic force $f_{\rm eq}$ invariant, so one may equivalently think of these forces as gradients of ${\cal F}$ or $\Phi$.

Similar to $s_{\rm rev}$, we find $s_{\rm interface}>0$ at the leading edge, because the current $\jbar_{\rm active}$  {is parallel to $\jbar_{\rm eq}$} there, and $s_{\rm interface}<0$ at the trailing edge where the current {is anti-parallel to $\jbar_{\rm eq}$}. In this way, activity leads to a cycle where free energy is constantly being fed into the system at the trailing edge and released at the leading edge; see discussion in Sec.~\ref{cycle} below. Another way to understand the excess dissipation at the leading interface is that the liquid phase is invading the vapour, and the liquid has the lower value of the relevant free-energy $\Phi$.  Hence the direction of interfacial motion is consistent with the equilibrium forces, while the opposite is true at the trailing edge. 


For repulsive active particles, which undergo a phase separation between dilute and dense regions, a reduction of EPR at interfaces has been reported~\cite{bebon_thermodynamics_2024}, consistently with results for field theories of scalar active matter~\cite{markovich_thermodynamics_2021}. In contrast with the travelling bands of aligning active paticles, interfaces of repulsive active particles do not move at constant speed. Therefore, active forces (namely, self-propulsion) of repulsive particles point towards dense regions, whereas passive forces (namely, steric repulsion) point towards dilute regions. Then, the active current (proportional to the local polarization) is in the direction opposite to the thermodynamic force, so the local correction to the bulk EPR always is always negative.


\subsection{Thermodynamic cycles in density-magnetization space}\label{cycle}

\begin{figure}
	\centering
	\includegraphics[width=.35\columnwidth]{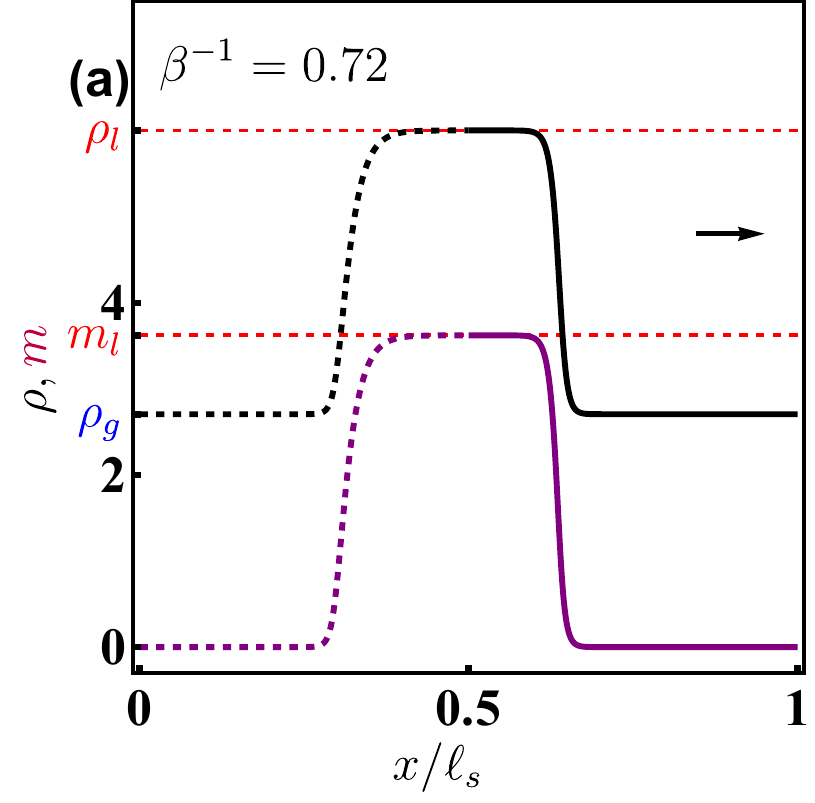}
	\hskip.1cm
	\includegraphics[width=.48\columnwidth]{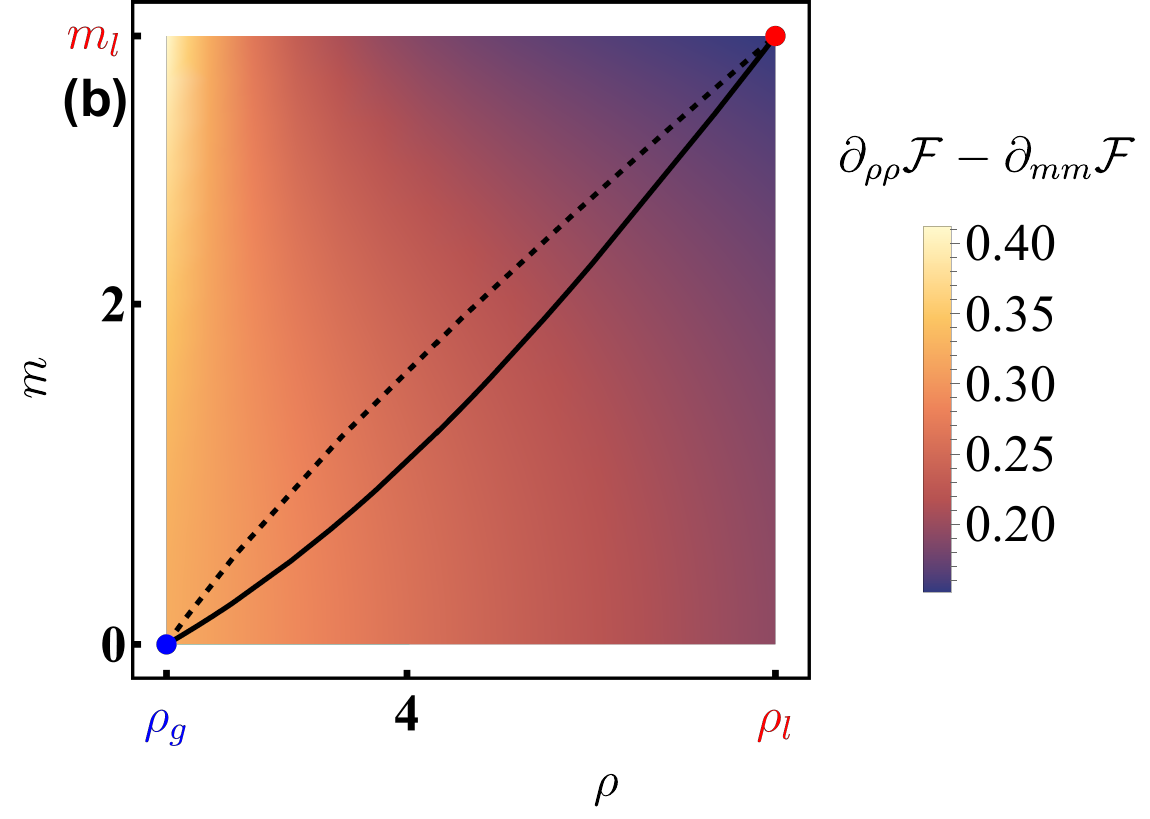}	
	\caption{(a)~Travelling band profile with the leading and trailing interfaces  shown in solid and dashed lines, respectively.
	(b)~Parametric representation of the same profile as a loop in the $(\rho,m)$ plane. Dots mark the binodal bulk phase values. The asymmetry of the two interfaces leads to a finite surface enclosed within the loop. The color map indicates the magnitude of the integrand defining the interfacial contribution to EPR [Eq.~\eqref{surface}]. Parameters: $\rho_0=2.625$, $\text{Pe=1}$.}
	\label{interface2}
\end{figure}

We now return to the total (spatially-integrated) EPR $\langle S_{\rm hydro} \rangle$ of the T.B. state. Recall from Eq.~\eqref{ent5} that this is generally less than its value in homogeneous states. To understand the size of this reduction, we substitute the expression of the thermodynamic flux [Eq.~\eqref{jbareqact}] and force [Eq.~\eqref{feqac}] into Eq.~\eqref{equ:sint}, to obtain
\begin{equation}\label{inter1}
	\begin{aligned}
		\langle S_{\rm hydro} \rangle - N {\rm Pe}^2 &= - \frac{L\text{Pe}}{\ell_s}\int_0^{\ell_s} dx\left(\rho\partial_x\frac{\partial\mathcal{F}}{\partial m}+m\partial_x\frac{\partial\mathcal{F}}{\partial \rho}\right)
		\\
		&= \frac{L\text{Pe}}{\ell_s}\int_0^{\ell_s} dx\left(\frac{\partial\mathcal{F}}{\partial m} \partial_x\rho + \frac{\partial\mathcal{F}}{\partial \rho} \partial_xm \right) ,
	\end{aligned}
\end{equation}
where the second line used integration by parts. In the T.B. state, the integral in Eq.~\eqref{inter1} depends on the shape of the travelling band. Now substitute in the travelling profile Eq.~\eqref{eq:travel} and cast the EPR [Eq.~\eqref{inter1}] as a closed loop integration: 
\begin{equation}\label{loop}
	\langle S_{\rm hydro} \rangle = N {\rm Pe}^2 + \frac{L\text{Pe}}{\ell_s}\oint_{\partial\Sigma}\left[\frac{\partial\cal F}{\partial m} d\rho+\frac{\partial\cal F}{\partial \rho}dm\right] ,
\end{equation}
where the integral is performed in the $\left(\rho,m\right)$ plane clockwise along the closed curve $\partial\Sigma$ defined parametrically by the band profiles $\left[\bar{\rho}(x),\bar{m}(x)\right]$ with $x\in\left[0,\ell_s\right)$ [Fig.~\ref{interface2}]. Applying Green's theorem, we write Eq.~\eqref{loop} as an integral over the surface $\Sigma$ enclosed within the closed curve $\partial\Sigma$:
\begin{equation}\label{surface}
	\langle S_{\rm hydro} \rangle = N {\rm Pe}^2 + \frac{L\text{Pe}}{\ell_s}\iint\limits_{\Sigma}d\rho dm\left(\frac{\partial^2\cal F}{\partial m^2} -\frac{\partial^2\cal F}{\partial \rho^2} \right) ,
\end{equation}
where we have used the convention of clockwise integration in Eq.~\eqref{loop}.

The integral in Eq.~\eqref{surface} depends on the asymmetry of the two interfaces. For example, if the travelling profile were left-right symmetric then the loop $\partial\Sigma$ would consist of two overlapping branches, so that the enclosed surface $\Sigma$ would vanish. In the language of~\ref{sec:interf-epr-local}, this corresponds to a situation where the extra free energy dissipated at the leading edge matches the free energy stored at the trailing edge, leading to a kind of reversible cycle that repeats as the band travels many times around the periodic boundaries. Such a balance of work input/output is present for $s_{\rm rev}$ but not generally for $s_{\rm interface}$.

The asymmetry of the travelling profile is shown in Fig.~\ref{interface2}, together with the cyclic (parametric) representation based on Eq.~\eqref{surface}.  Consistent with this, the bound $\langle S_{\rm hydro}\rangle\leq N {\rm Pe}^2$ [Eq.~\eqref{ent5}] is not saturated in practice, which also shows that the energy exchanges are not symmetric at the leading/trailing edge.  The representation in Eq.~\eqref{surface} provides a route to evaluating EPR for an arbitrary profile. In particular, it sets the stage for some perturbative approaches based on asymptotically determining the corresponding cycle in $(\rho,m)$ space, which we discuss next.

\subsection{Scalings of EPR close to equilibrium}\label{eprlow}

\subsubsection{Interfacial EPR.}

\begin{figure}
	\centering
	\includegraphics[width=.295\columnwidth]{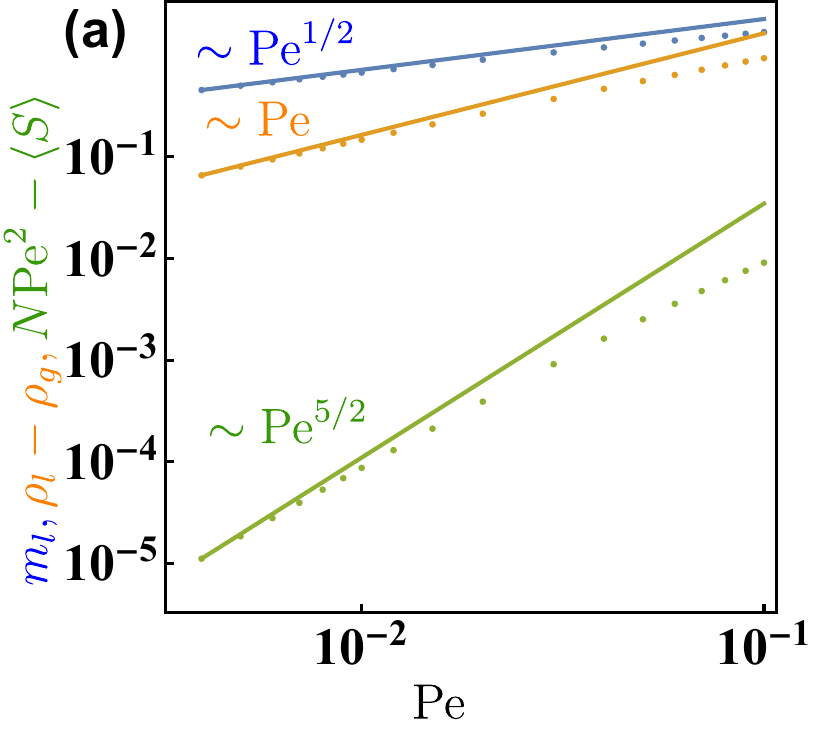}	
	\includegraphics[width=.3\columnwidth]{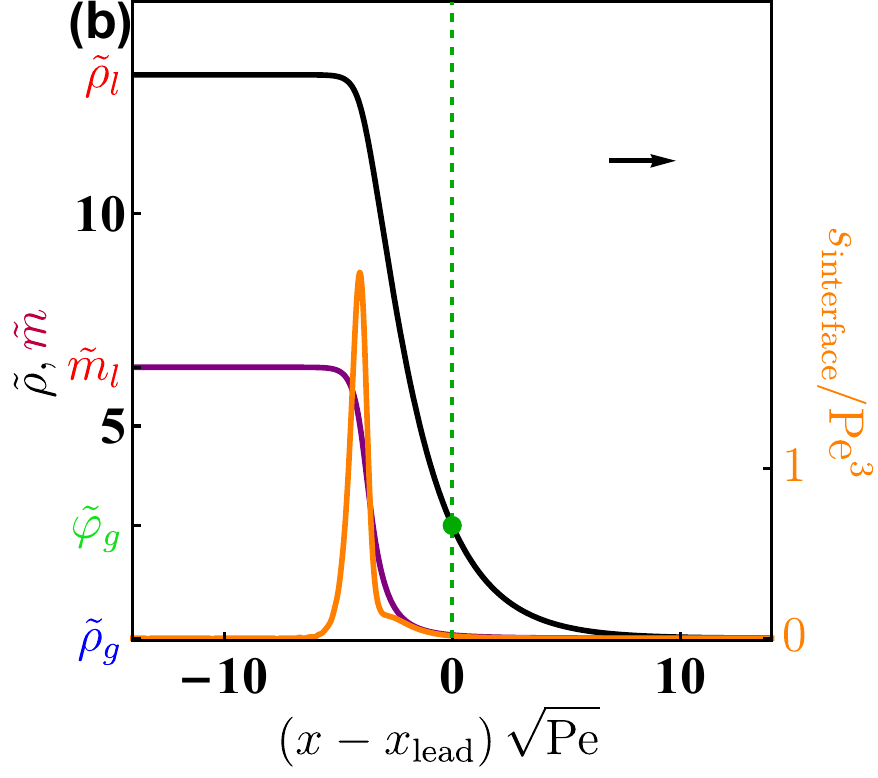}
	\includegraphics[width=.37\columnwidth]{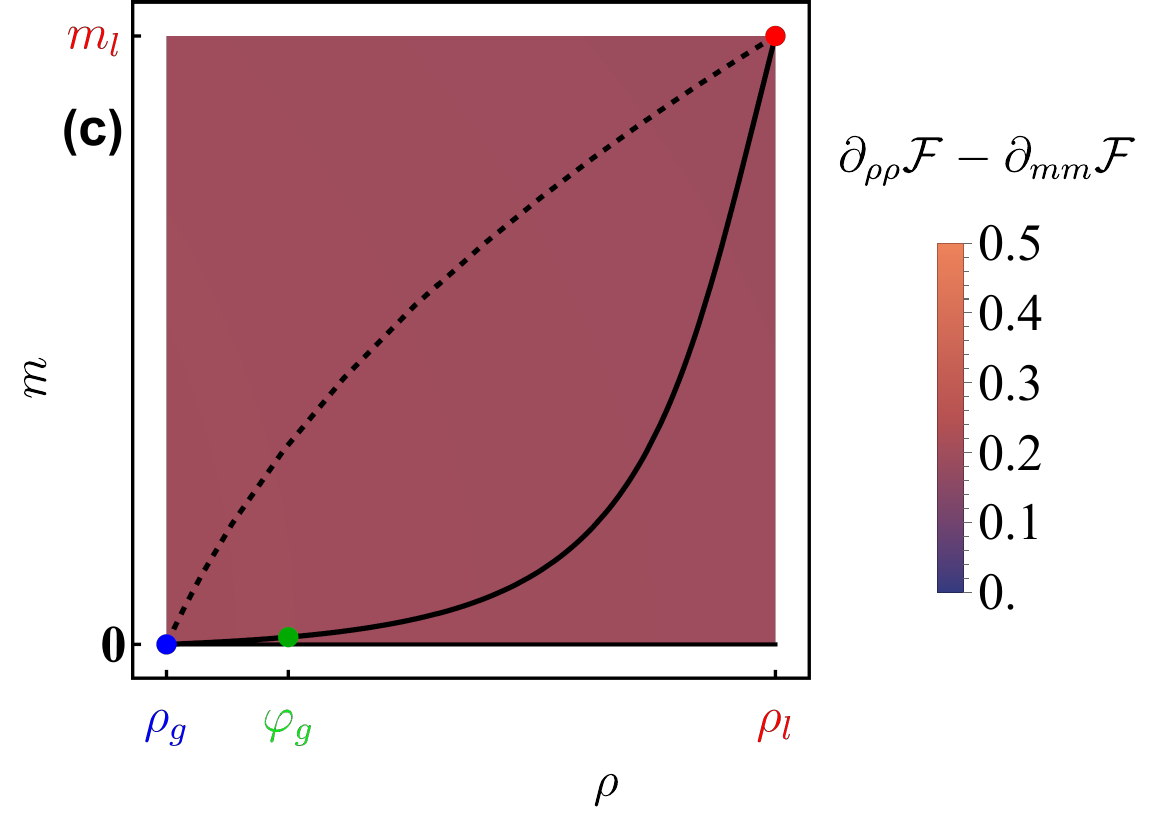}	
	\caption{The interfacial EPR at low $\text{Pe}=0.02$.  
		(a)~Same as in Fig.~\ref{narrow}(c) with additional data from the interfacial EPR [Eq.~\eqref{inter1}] evaluated from numerical solutions for the travelling profiles and where we set $L/\ell_s=1$. The straight green line has the slope $-5/2$ in agreement with analytical predictions [Eq.~\eqref{surface2}].
		 (b)~Same as in Fig.~\ref{lowpe} with additional data for the interfacial EPR density [Eq.~\eqref{inter1}] which is restricted to the back section of the leading interface where the density is above the gaseous spinodal $\rho>\varphi_g$.
		 (c)~Surface enclosed by the travelling wave profiles in the $\rho,m$ plane which gives the EPR at low $\text{Pe}$ [Eq.~\eqref{surface2}]. The color map shows the integrand \eqref{surface}, whose value depends only weakly on $(\rho,m)$ throughout the enclosed region $\Sigma${, in contrast to Fig~\ref{interface2}(b)}.
		 }
	\label{lowpeepr}
\end{figure}

We now analyse the interfacial contribution at small $\Pe$. Recall from Eq.~\eqref{expand0} that the differences of density and magnetisation between the dense and dilute phases scale as $\Pe$ and $\Pe^{1/2}$ respectively. It follows that the size of the density-magnetization loop $\Sigma$, which defines the interfacial EPR [Eq.~\eqref{surface}], shrinks to zero as Pe decreases. In fact, $\Sigma$ approaches a limiting shape, for which
\begin{equation}
\iint_\Sigma d\rho dm \sim \Pe^{3/2} .
\label{equ:Sigma-scaling}
\end{equation}
As a representative point inside this loop, we take $(\rho,m)=(\varphi_g,0)$, where $\varphi_g=\beta/\left(\beta-1\right)$ is the density on the gas spinodal, defined so that $\partial_{mm}{\cal F}=0$ and $\partial_{\rho\rho}{\cal F}=1/\varphi_g$ [Eqs.~(\ref{sping},\ref{eq:sping-partials})]. Since the loop $\Sigma$ is small and ${\cal F}$ is smooth, we deduce that the integrand of the interfacial EPR contribution [Eq.~\eqref{surface}] reads 
\begin{equation}
	\frac{\partial^2\cal F}{\partial \rho^2}-\frac{\partial^2\cal F}{\partial m^2} \approx \frac{\beta-1}{\beta} + \mathcal O(\text{Pe}^{1/2})
	\label{equ:curv}
\end{equation}
everywhere in $\Sigma$. The correction is given by the maximal linear extent of $\Sigma$. Hence, using Eqs.~(\ref{surface}, \ref{equ:Sigma-scaling}), the interfacial contribution to the total EPR is given by
\begin{equation}\label{surface2}
	N {\rm Pe}^2-\langle S_{\rm hydro} \rangle \sim \frac{L\text{Pe}}{\ell_s\varphi_g} \iint_\Sigma d\rho dm \sim \Pe^{5/2} .
\end{equation}
{which can be compared with the bulk contribution's scaling $\sim\text{Pe}^2$. }Numerical results in Fig.~\ref{lowpeepr}(a) confirm {the} scaling \eqref{surface2}.

To explore this behaviour in more detail, Fig.~\ref{lowpeepr}(b) shows how the local interfacial EPR depends on position.  It is localised at the interfaces with positive and negative contributions similar to Fig.~\ref{interface}, recall the discussion of Sec.~\ref{sec:interf-epr-local}. Fig.~\ref{lowpeepr}(c) shows the parametric representation of the density profile; results in \ref{ap.lowpe} show that the small-Pe behaviour of this plot can be obtained from the phase portrait of a dynamical system [Fig.~\ref{dynsys}]. In particular, the green dot in Fig.~\ref{lowpeepr}(c) indicates the point $(\rho,m)=(\varphi_g,0)$ where the boundary of the limiting shape is singular, because it follows the line $m=0$ between this point and $(\rho_g,0)$. Interestingly, the {scaled} local interfacial EPR $s_{\rm interface}/\text{Pe}^3$ vanishes in this part of the interface. 
Physically, this region with $m\approx 0$ and $\rho_g < \rho < \varphi_g$ also has {$\jbar_{{\rm eq},m}=0$ [Eq.~\eqref{jbareqact}]} and hence $s_{\rm interface}=0$ also [Eq.~\eqref{decomp}]. In this region, the positive velocity of the band appears because the equilibrium force drives a (diffusive) density current to the right, down the density gradient. Elsewhere in the leading edge, the active forces are finite, and sustain the travelling band.

\subsubsection{EPR contributions from individual particle types.}

The scaling in Eq.~\eqref{expand0} has another unexpected outcome on the interfacial EPR that we briefly describe. It follows from considering the separate contribution to the EPR from each particle type:
\begin{equation}
	s_{\rm bulk}=s^+_{\rm bulk} + s^-_{\rm bulk} , \qquad  s_{\rm interface}=s^+_{\rm interface} + s^-_{\rm interface} 
\end{equation} 
with currents and forces for $\pm$ particles defined as
\begin{equation}
	s^\pm_{\rm bulk} = \jbar_{\rm active}^\pm f_{\rm active}^\pm , \qquad  s^\pm_{\rm interface} = \frac12 ( \jbar_{{\rm eq},\rho} \pm \jbar_{{\rm eq},m} ) f_{\rm active}^\pm ,
\end{equation}
where $ \jbar_{\rm active}^\pm = \Pe( m\pm \rho)/2$ and $f_{\rm active}^\pm = \pm \Pe$ [Eqs.~(\ref{feqac},\ref{jbareqact})].

As we have seen, in both the H. and C.M. phases, the only contribution to the EPR is coming from the bulk piece [Eq.~\eqref{bulk}] where the contribution per particle is constant and equal for both particle types:
\begin{equation}\label{bulksym}
	\frac{s^+_\text{bulk}}{\rho_+} = \frac{s^-_\text{bulk}}{\rho_-} = \text{Pe}^2 ,
\end{equation} 
where $\rho_{\pm}=(\rho\pm m)/2$ are the $+$ and $-$ particle number density fields. However, in the T.B. phase, where an extra interfacial contribution is picked up, this apparent symmetry is in general broken with different contributions form the two particle types. The interfacial contribution from the two types is given by
\begin{equation}\label{pm}
	s^\pm_\text{interface} = \mp\text{Pe}\rho_\pm\partial_x\frac{\partial\mathcal{F}}{\partial\rho_\pm} ,
\end{equation}
and correspondingly
\begin{equation}
	\frac{\partial\mathcal{F}}{\partial \rho_{\pm}}=\frac{\partial\mathcal{F}}{\partial \rho}\pm\frac{\partial\mathcal{F}}{\partial m} .
\end{equation}
Then, Eq.~\eqref{expand0} implies that
\begin{equation}
	\frac{\partial\mathcal{F}}{\partial m} \sim m \sim {\Pe}^{1/2} \quad,\quad \frac{\partial\mathcal{F}}{\partial \rho} = \mathcal O\left(1\right) ,
\end{equation}
so the two species feel equal free-energy gradients $\partial\mathcal{F}/\partial\rho^{+} \approx \partial\mathcal{F}/\partial\rho^{-}$. Then, Eq.~\eqref{pm} yields
\begin{equation}\label{asym}
	\frac{s^+_\text{interface}}{\rho_+} = - \frac{s^-_\text{interface}}{\rho_-} + \mathcal{O}(\Pe^{3}) .
\end{equation}
Therefore, the interfacial EPR per particle is equal and opposite for the two species. The individual terms are $O(\Pe^{5/2})$, as in Eq.~\eqref{surface2}.


\section{Discussion}\label{disc}

We {have studied} the entropy production rate measuring the energy dissipation in a flocking lattice model where self-propulsion is the only source of external work. Analytical progress is possible here thanks to the diffusive scaling of the microscopic rates which enables exact coarse-graining. This leads to equality between the IEPR at hydrodynamic scales and the microscopic dissipation. Despite these idealised modelling assumptions, we expect that many qualitative characteristics of the EPR found here to be generic in other themodynamically consistent models of active matter. Indeed, although our findings are formulated in terms of coarse-grained fields, they are corroborated by microscopic arguments that do not rely on the diffusive scaling of the rates. In particular, our methods for analysing EPR could be deployed in other models of active matter, such as non-reciprocal Ising models~\cite{Rieger2023, avni2025a, avni2025b, mohite2025} and pulsating active particles~\cite{manacorda2024}, provided that these models satisfy the condition of local detailed balance at the basis of thermodynamic consistency.

\subsection{Comparison with other models}

Let us summarize our findings in comparison to previous models of flocking, where dissipation exhibits a distinct behavior due to a different source of irreversibility.

In discrete flocking models, like the active Ising model (AIM)~\cite{solon_revisiting_2013,solon_flocking_2015} and its recent extensions~\cite{kourbane-houssene_exact_2018-1,scandolo_active_2023}, the dynamics of positions is decoupled from the Hamiltonian controlling the dynamics of orientations. This decoupling can also be regarded as a temperature mismatch between the two thermal baths that are in contact with the position and orientation degrees of freedom~\cite{solon_flocking_2015, agranov_thermodynamically_2024}. This situation creates another source of irreversibility in addition to the one associated with self propulsion. In particular, dissipation remains finite at vanishing self propulsion~\cite{solon_flocking_2015, yu_energy_2022}. In continuous symmetry flocks, like the Vicsek model~\cite{vicsek_novel_1995,toner_long-range_1995} and its variants~\cite{ferretti_signatures_2022,ferretti_out_2024}, there is a similar thermodynamic inconsistency between the two types of degrees of freedom. Even if the alignment dynamics are described by a Hamiltonian and respect detailed balance~\cite{ferretti_signatures_2022, ferretti_out_2024}, the positional updates of the particles do not feel this Hamiltonian, so the model is not thermodynamically consistent, see also~\cite{prawar_dadhichi_origins_2018}.

This extra source of irreversibility brings qualitative differences in the nature of dissipation compared to our findings here. First, since dissipation here is directly linked to particle's persistent motion, and since the latter is unchanged between the homogeneous disordered and homogeneous flocked states, then so is the dissipation. This is captured by our explicit expression for bulk dissipation [Eq.~\eqref{bulk}]. This is contrary to the findings in previous models that reported on a cusped dissipation peak at the flocking transition~\cite{yu_energy_2022, ferretti_signatures_2022, ferretti_out_2024}. The latter originates from spontaneous alignment with a corresponding singular change in the rate at which the thermodynamic mismatch between orientation and translation dissipates energy. It does not represent a change in the energetic cost of propelling particles, and in particular the cusped peak was shown to survives in the zero self propulsion limit~\cite{yu_energy_2022}.

\subsection{Spatial decomposition of EPR}

Our second finding concerns the spatial decomposition of EPR, showing a strong spatial modulation of EPR at phase boundaries within the travelling band state. A similar behavior was recently reported for continuous flocks using a deep learning approach~\cite{Boffi_2024}. Within our analytically tractable field theory, we link these modulations to particles being driven either up or down free energy gradients. Particles entering the travelling band from the {leading} edge gain free energy, while those that leave the {trailing} edge lose it. The overall effect of these processes is to reduce the EPR with respect to the value attained in homogeneous states. We propose a cyclic description in the ($\rho,m)$ plane to rationalise this effect. Moreover, our analysis at small $\text{Pe}$ reveals a singular scaling of dissipation, and provides further insights into the mechanism sustaining the phase-separated travelling bands.

We expect spatial modulations of the local EPR to be generic in thermodynamically consistent flocking systems, since spatial modulations in the free energy generically modulate the rate of particles displacement in such models. It would be interesting to examine how our analysis extends to other phase behaviors featuring both reversed and counter-propagating bands~\cite{agranov_thermodynamically_2024}. 


\section*{Acknowledgements}

The authors acknowledge useful discussions with Chiu Fan Lee, Massimiliano Esposito, Gianmaria Falasco, Karel Proesmans, and Atul Tanaji Mohite. This research was funded in part by the Luxembourg National Research Fund (FNR), grant reference 14389168, and grant no. NSF PHY-2309135 to the Kavli Institute for Theoretical Physics (KITP).


\appendix

\section{Stochastic hydrodynamics}\label{ap.fluc}

The microscopic dynamics described in Sec.~\ref{model} consists of biased diffusive hops along with reaction dynamics. Fluctuating hydrodynamics of this type of dynamics was derived in~\cite{bodineau_current_2010-1}, and has been implemented in other active lattice gas models~\cite{kourbane-houssene_exact_2018-1, agranov_exact_2021, agranov_entropy_2022, agranov_macroscopic_2022}. To derive the the flux noise terms, we start with the microscopic rates of a particle to hop from site $i$ to its neighboring sites $i\pm1$. As was shown in~\cite{agranov_thermodynamically_2024}, the corresponding probability jump rates can be written for both particle types as 
\begin{equation}\label{u}
	\text{rate}(i\rightarrow i\pm1) =D_0+\mathcal O(L^{-1}) .
\end{equation}
Crucially, the leading order terms are symmetric hops with fixed rate $D_0$, which guarantees that the measure is controlled by the simple symmetric hop rules corresponding to noninteracting random walkers~\cite{bodineau_current_2010-1}. At the hydrodynamic level, one can then write down the corresponding fluctuations of the fluxes $J_{\pm}=\left(J_{\rho}\pm J_{m}\right)/2$ as~\cite{bertini_macroscopic_2015-1}
\begin{equation}\label{gnoise}
	J_{\pm}\left(x,t\right)-\bar{J}_{\pm}\left(x,t\right)=\sqrt{\frac{\ell_s}{L}}\eta_{\pm} ,
\end{equation}
where $\bar J_\pm$ refers to average values [Eq.~\eqref{d2}], and $(\eta_+,\eta_-)$ are some uncorrelated Gaussian white noises with zero mean and correlations given by
\begin{equation}
	\langle\eta_{\pm}\left(x,t\right)\eta_{\pm}\left(x',t'\right)\rangle=2\rho_{\pm}\delta(x-x')\delta(t-t'),
\end{equation}
in terms of the coarse-grained units of space and time [Sec.~\ref{hyd}]. The corresponding form of the diffusive Lagrangian ${\cal L}_J$ [Eq.~\eqref{fluxrate}] directly follows.

To account for fluctuations in the tumbling rate we follow the large-deviation formalism for reactive lattice gases~\cite{jona-lasinio_large_1993, bodineau_current_2010-1, agranov_macroscopic_2022}. We consider a mesoscopic interval of space $[x,x+\Delta x]$ and time $[t,t+\Delta t]$, for which the density $\rho_{\pm}\left(x,t\right)$ can be regarded as locally constant. The corresponding number of $+$ particles within this interval reads $\rho^+L\Delta x/\ell_s \sim L^\delta\gg1$, each of them tumbling with probability $r^+\Delta t$ during $\Delta t$ , where
\begin{equation}
	r^+ = e^{\beta\frac{\delta H}{\delta m}} , 
\end{equation}
is the probability tumble rate on macroscopic time units~\cite{agranov_thermodynamically_2024}. The total number of tumbling events of $+$ into $-$ obeys a Poisson distribution: $\mathcal K_+\sim \text{Pois}(L\Delta x\Delta t\rho^+r^+/\ell_s)$. Then, the tumbling rate density $K_+=\frac{\mathcal K_+\ell_s}{L\Delta x\Delta t}$ follows a large deviation principle given by
\begin{equation}
	-\ln P\left(K_+\right)\simeq \frac{L}{\ell_s}\Delta x\Delta t\,\psi_{\rho^+r^+}(K_+) ,
\end{equation}   
with
\begin{equation}
	\psi_{\rho^+r^+}(K_+)=K_+\log\left(\frac{K_+}{\rho^+r^+}\right)-K_++\rho^+r^+ .
\end{equation}
Similarly, the corresponding tumble rate density of the opposite tumbling reaction ($-\to+$), denoted by $ K_-$, follows the large deviation principle given by
\begin{equation}
	-\ln P\left(\mathcal K_-\right)\simeq \frac{L}{\ell_s}\Delta x\Delta t\,\psi_{ \rho^-r^-}(K_-) ,
\end{equation}
with
\begin{equation}
	r^- = e^{-\beta\frac{\delta H}{\delta m}} . 
\end{equation}
Lastly, following the contraction principle~\cite{touchette_large_2009-2}, the distribution of the tumbling difference $ K= K_+- K_-$ is described by 
the large deviation function
\begin{equation}\label{phi}
	-\ln P\left( K\right)\simeq \frac{L}{\ell_s}\Delta x\Delta t \, \mathcal L _K(K) ,
\end{equation}
where $\mathcal L _K$ is found by minimizing the combined probability cost of the previous two processes under the constraint of tumbling rate difference:
\begin{equation}
	\mathcal L_K=\inf_{K_+}\left[\psi_{ r^+\rho^+}(K_+)+\psi_{ r^-\rho^-}(K_+-K)\right]\label{phi1}
\end{equation}
which, after some algebra, can be shown to coincide with the expression in Eq.~\eqref{phi12}, where the mean tumbling $\bar{K}$ and the mobility $M$ read
\begin{equation}
	\bar{K}= r^+\rho^+-r^-\rho^-\quad,\quad	M=2\sqrt{r^+\rho^+r^-\rho^-}.
\end{equation}
After some algebra, these can be shown to coincide with the expressions in Eqs~\eqref{d2} and~\eqref{mob}. Lastly, summing up the contribution from all the mesoscopic compartments, we arrive at the action in Eq.~\eqref{act} given in terms of the tumbling Lagrangian in Eq.~\eqref{phi1}.


\section{Travelling bands close to equilibrium}\label{ap.lowpe}

In this Appendix, we derive some analytical results regarding the behavior close to equilibrium at small $\Pe$.

\subsection{Coexistence between polar bands and apolar background}

We now derive the scalings for the binodal gaps, magnetisation, and velocity of travelling bands (T.B.) as given Eq.~\eqref{expand0}. The solid red and blue lines in Fig.~\ref{schem}(b) are the binodals, and the region between these lines marks the miscibility gap, within which one observes macroscopically inhomogeneous (T.B.) states. The dashed lines are the spinodals which mark the limits of stability of the ordered and disordered homogeneous phases, as derived in~\cite{agranov_thermodynamically_2024}. We show in this section that the difference of the spinodals is $O(\Pe)$, and Fig.~\ref{schem}(c) confirms that the same scaling holds for the binodals at sufficiently small Pe.

We describe the spinodal lines by temperature-dependent densities: $\varphi_l(\beta)$ and $\varphi_g(\beta)$. From~\cite{agranov_thermodynamically_2024}, the spinodal $\varphi_g$ of the disordered gas phase [dashed blue lines in Figs.~\ref{schem}(b) and~\ref{narrow}(b)] is given by  $\beta g(\varphi_g) = 1 $ with $g$ given by Eq.~\eqref{f0}, so 
\begin{equation}\label{sping}
	\varphi_g(\beta) = \frac{\beta}{\beta-1}  .
\end{equation}
This spinodal instability is the onset of spontaneous magnetisation~\cite{agranov_thermodynamically_2024}, so that for $\rho=\varphi_g(\beta)$, we obtain 
\begin{equation}
\label{eq:sping-partials}
\partial_{mm}\mathcal F=0, \qquad \partial_{\rho\rho}\mathcal F = 1/\varphi_g .
\end{equation}
The spinodal $\varphi_l$ of the ordered liquid phase [dashed red lines in Figs.~\ref{schem}(b) and~\ref{narrow}(b)] is given implicitly by the relation 
\begin{equation}\label{instp}
	\text{Det}\left[\text{Hess}\left({\cal F}\right)\right]\vert_{\left(\rho=\varphi_l,m=m_l\right)}=\frac{\text{Pe}^2}{2\rho M(\rho,m)}\left[\left(\frac{\partial_{\rho m}\mathcal F}{\partial_{m m}\mathcal F}\right)^2-1\right]_{\left(\rho=\varphi_l,m=m_l\right)} ,
\end{equation}
with $\text{Det}\left[\text{Hess}\left({\cal F}\right)\right]$ the Hessian determinant of $\mathcal F\left(\rho,m\right)$, the mobility $M$ given in Eq.~\eqref{mob}, and the bulk magnetization of the liquid $m_l=m_0(\varphi_l,\beta)$ {is} defined by
\begin{equation}\label{mstar}
	m_0(\rho,\beta) = \rho z(\rho,\beta) ,
	\quad
	\frac{\operatorname{tanh}^{-1} z}{z}=\beta g(\rho) ,
\end{equation}
where the second equality defines the dependence of $z$ on $\rho$ and $\beta$.

{To establish the small $\text{Pe}$ scaling for the liquid spinodal we define 
$\epsilon\equiv\varphi_l\left(\beta\right)-\varphi_g\left(\beta\right)$ 
 which by \eqref{sping} is equivalent to
\begin{equation}
\label{varphipe}
\varphi_l = \frac{\beta}{\beta-1} + \epsilon
\end{equation}
The two spinodals both approach the critical line as $\Pe\to0$ so $\epsilon$ is small in this limit.  Evaluating \eqref{mstar} at $\rho=\varphi_l$ and using this fact, one finds that $z$ is also small, leading to an expression for the magnetization at leading order in $\epsilon$,
\begin{eqnarray}\label{mlpe}
	m_l\simeq\sqrt{3\beta\epsilon}.
	\end{eqnarray}
Plugging \eqref{varphipe} and \eqref{mlpe} into \eqref{instp}, and keeping only leading order terms in $\epsilon$, we finally arrive at \footnote{The quantity under the square root in \eqref{varphilpe} is positive for any temperature above the tricritical temperature, $\beta^{-1}>\beta^{-1}_{\text{tri}}=3-\sqrt{6}$ }
\begin{eqnarray}\label{varphilpe}
\varphi_l\left(\beta\right)\simeq\varphi_g\left(\beta\right)+\text{Pe}\frac{\beta}{2\left(\beta-1\right)\sqrt{4\beta-2\beta^2-\frac{2}{3}}}
	\end{eqnarray}
which, when plugged into~\eqref{mlpe} also implies $m_l=\mathcal O(\text{Pe}^{1/2})$.} 
Combined with the expression of the band velocity [Eq.~\eqref{eq:VV}], we have thus fully established the scaling laws in Eq.~\eqref{expand0}.


\subsection{Scaling of the T.B. profile}

In this section, we establish the singularity for the profile of the travelling band at low $\text{Pe}$ as described in Sec.~\ref{lowpelimit}. It is associated with a transition in the scaling for $m$ along the leading interface: in the back section of this interface $m$ scales like $\text{Pe}^{1/2}$, whereas $m$ in the leading section scales like $\text{Pe}^{3/2}$. To examine such a transition, we derive a set of ordinary differential equations (ODEs) that describe the interfacial profile, and expand these ODEs at low $\text{Pe}$.

The T.B. profile is found by plugging $\rho=\bar{\rho}(x-Vt)$ and $m=\bar{m}(x-Vt)$ into Eqs.~(\ref{d1}-\ref{d2}) which results in the coupled second order ODEs:
\begin{equation}\label{d3}
	\begin{bmatrix}
		-V\bar{\rho}'\\
		-V\bar{m}'
	\end{bmatrix}=\left(\frac{1}{2}\mathbb C(\bar{\rho},\bar{m})\text{Hess}\left({\cal F}\right)\begin{bmatrix}
		\bar{\rho}'\\
		\bar{m}'
	\end{bmatrix}\right)'-\text{Pe}\begin{bmatrix}
		\bar{m}'\\
		\bar{\rho}'
	\end{bmatrix}-\begin{bmatrix}
		0\\
		2M(\bar{\rho},\bar{m}) \sinh\left(\frac{\partial F}{\partial m}\right)
	\end{bmatrix} ,
\end{equation} 
with primes denoting derivatives with respect to the argument. Since the $\rho$ dynamics is conservative, the first row of \eqref{d3} can be integrated once:
\begin{equation}\label{rhoode}
	\left(\bar\rho \partial_{\rho \rho}{\cal F} + m \partial_{\rho m}{\cal F}\right)\bar{\rho}'+\left(\bar\rho \partial_{\rho m}{\cal F} + \bar m \partial_{m m} {\cal F}\right)\bar{m}'+V\bar{\rho}-\text{Pe}\,\bar{m} = C .
\end{equation}
The integration constant $C$ can be evaluated in either of the bulk phases to yield
\begin{equation}\label{c}
	C = V\rho_g = V\rho_l-\text{Pe}\, m_l .
\end{equation}
Then, using the scalings in Eq.~\eqref{expand0}, we deduce
\begin{equation}\label{v}
	V=\text{Pe}\frac{m_l}{\rho_l-\rho_g}=\mathcal O(\text{Pe}^{1/2}).
\end{equation}
which sets the scaling of the velocity of the travelling band.

\begin{figure}
	\centering
	\includegraphics[width=.35\columnwidth]{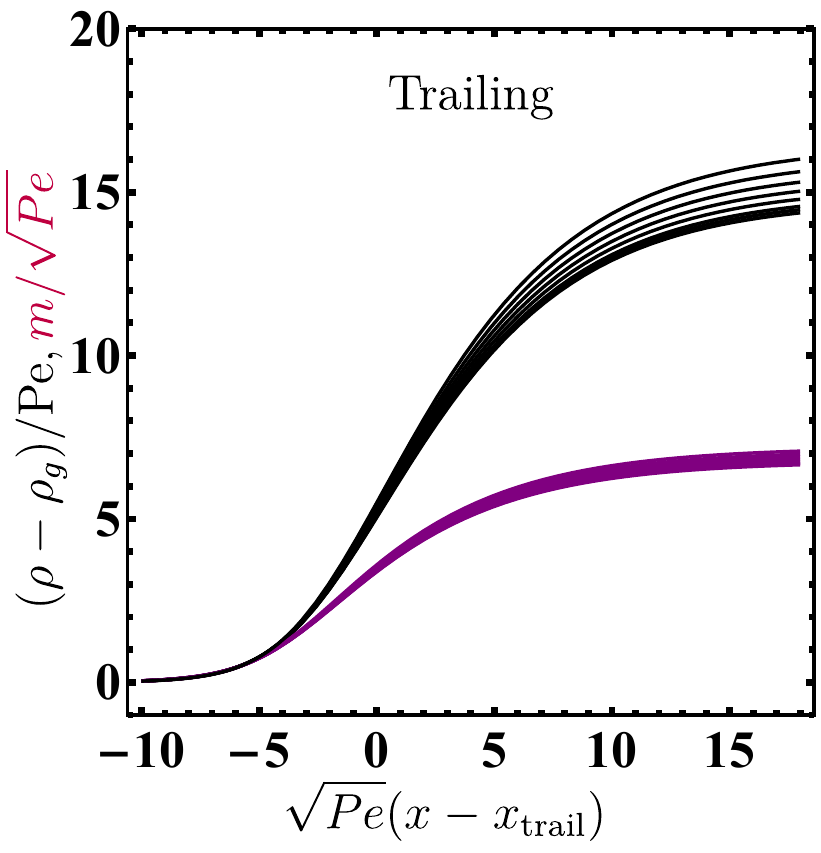}	
		\includegraphics[width=.35\columnwidth]{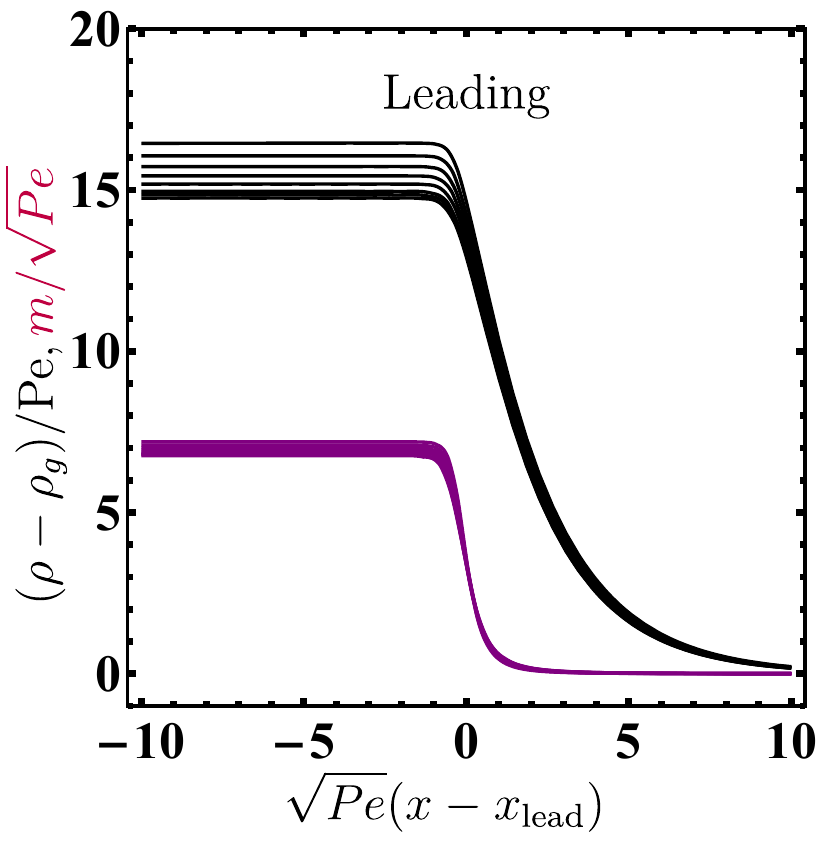}	
	\caption{Validation of the scaling for miscibility gap [Eq.~\eqref{xscale}] and	interfacial width [Eq.~\eqref{expand0}] for the travelling band interfaces at low $\text{Pe}$. The plots show an overlay of the travelling bands for seven values of $\text{Pe}=(0.01,0.009,0.008,\dots,0.004)$, decreasing from top to bottom. Here $x_{\text{trail}}$ and $x_{\text{lead}}$ are the locations where the corresponding magnetization profile reaches half of its height.}
	\label{scale}
\end{figure}

The partial derivatives of the free energy [Eq.~\eqref{free}] can be computed explicitly. Plugging these and the value of $C$ [Eq.~\eqref{c}] into Eq.~\eqref{rhoode}, we arrive at 
\begin{equation}\label{comp}
	\left(1+\frac{\beta\bar{m}^2}{\bar{\rho}^3}\right)\bar{\rho}'-\frac{\beta\bar{m}}{\bar{\rho}^2}\bar{m}'=V\left(\rho_g-\bar{\rho}\right)+\text{Pe}\,\bar{m} .
\end{equation}
From the scalings of the miscibility gap [Eq.~\eqref{expand0}] and the band velocity [Eq.~\eqref{v}], we deduce that the right-hand side of Eq.~\eqref{comp} vanishes in the bulk phases, and is $\mathcal O(\text{Pe}^{3/2})$ in the interfacial regions.  To ensure that leading order contribution from the left-hand side of Eq.~\eqref{comp} is also $\mathcal O(\text{Pe}^{3/2})$, we consider the scaled coordinate, relative to the interfacial position $x_{\rm interface}$:
\begin{equation}\label{xscale}
	\tilde x= (x - x_{\rm interface}) \sqrt{\Pe} .
\end{equation} 
Since the density and magnetisation differences between the phases are also vanishing at small $\Pe$, we also expand them about the gas spinodal $(\rho=\varphi_g,m=0)$ which yields 
\begin{equation}\label{expand1}
	\begin{aligned}
		\bar{\rho}{(\tilde x)} &= \varphi_g+\text{Pe}\,q_1(\tilde x)+\mathcal O\left(\text{Pe}^2\right) ,
		\\
		\bar{m}(\tilde x)&=\text{Pe}^{1/2}m_1(\tilde x)+\mathcal O\left(\text{Pe}^{3/2}\right) ,
		\\
		V &= \text{Pe}^{1/2}\,V_1+\mathcal O\left(\text{Pe}^{3/2}\right) .
	\end{aligned}
\end{equation}
A numerical verification of this scaling at the interfaces is shown in Fig.~\ref{scale}, this also implies \eqref{equ:travel-scale-main} for the vicinity of the leading edge. Note that $\rho_g - \varphi_g = {\cal O}(\Pe)$ so $\tilde \rho_{\rm lead}$ differs from the corresponding piece of $q_1$ by an additive constant.


\subsection{Singular form of TB profile}

We now explain the singular behaviour of $\tilde{m}_{\rm lead}$ that was discussed in Sec.~\ref{lowpelimit}.
Substituting \eqref{expand1} into the equation for $m$ [Eq.~\eqref{d3}], we arrive at two ODEs:
\begin{equation}\label{system1}
\begin{aligned}
	\dot{m}_1 &= -\frac{2\beta}{\varphi_g^2V_1}m_1\left(q_1-\frac{m_1^2}{3\beta}\right) ,
	\\
	\dot{q_1} &= V_1\left(c-q_1\right) + m_1-\frac{2\beta^2}{\varphi_g^4V_1}m_1^2\left(q_1-\frac{m_1^2}{3\beta }\right) ,
\end{aligned}
\end{equation}
where the dots indicate derivatives with respect to the scaled position $\tilde{x}$ [Eq.~\eqref{xscale}]{. This notation is used here to connect with the dynamical systems picture coming up next.} The constants
\begin{equation}\label{cval}
	c=\frac{\rho_g-\varphi_g}{\text{Pe}} ,
	\quad
	V_1 = \frac{ m_l \sqrt{\Pe}  }{ \rho_l - \rho_g }
\end{equation}
are fixed by Eq.~\eqref{expand0} to have values of order unity. We assume in the following that the liquid has positive magnetisation, leading to $V_1>0$. The other case can be dealt with similarly. In general, $c<0$ since the binodal density of the gas is lower than the corresponding spinodal.

\begin{figure}
	\centering
	\includegraphics[width=.38\columnwidth]{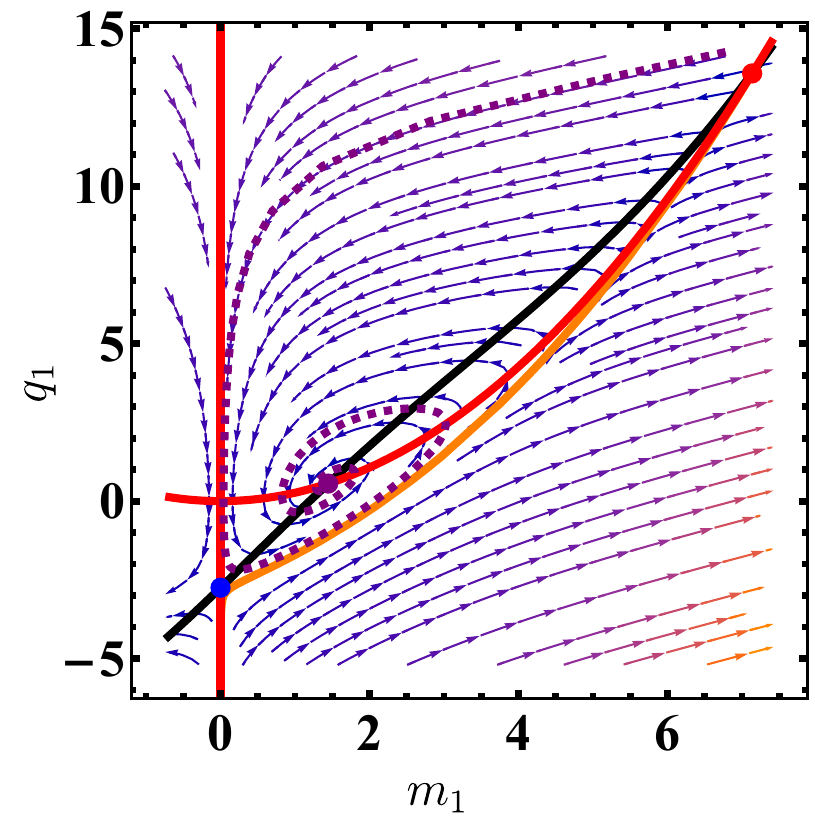}	
	\includegraphics[width=.38\columnwidth]{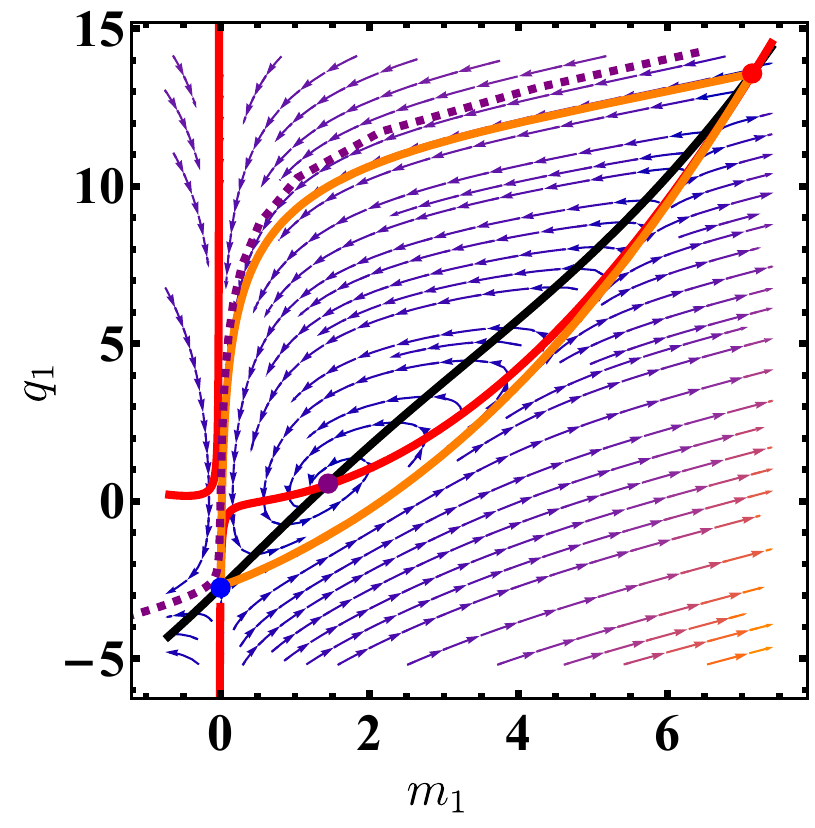}	
	\caption{{(a) Phase portrait for $(q_1,m_1)$ under the dynamics of Eq.~\eqref{system1}.  (b) Similar phase portrait using Eq.~\eqref{system1b}.}  Black and red solid lines are the $\dot q_1=0$ and $\dot m_1=0$ nullclines, respectively: they intersect at the two fixed points marked by the red and blue dots [Eq.~\eqref{fixed}] corresponding to the gaseous and liquid phases, respectively. The third fixed point marked by the purple dot is a stable spiral, and the orange solid lines denote heteroclines. Trajectories initiating above the heterocline in (a) always end up in the spiral, whereas trajectories initiating above the upper heterocline in (b) evade the spiral. The values of $c\simeq-2.75$ and $V_1\simeq0.43$ were extracted from the numerical solutions of the hydrodynamics in Eqs.~(\ref{d1}-\ref{d2}) at $\text{Pe}=0.004$ using the relations in Eqs.~\eqref{v} and~\eqref{cval}. 
	}
	\label{dynsys}
\end{figure}

The coupled ODEs in Eq.~\eqref{system1} can be interpreted as a dynamical system in the phase space $(q_1,m_1)$ [Fig.~\ref{dynsys}(a)]: the time variable in the dynamical system is the (scaled) position co-ordinate for the travelling profile.  The two independent parameters $V_1$ and $c$ encode the locations of fixed points in the dynamical system, and set the topology of the phase portrait. The ODEs in Eq.~\eqref{system1} have three fixed points, two of them are saddles that correspond to the bulk phases of the travelling bands:
\begin{equation}\label{fixed}
	\begin{aligned}
	\left[q_1^g, m_1^g\right] &= [c,0] ,
	\\
	[q_1^l, m_1^l] 	&= \Big[c+ 3\beta/(2V_1^2)+c\sqrt{[1+3\beta/(2c^2 V_1^2)]^2-1},\sqrt{3\beta q_1^l}\Big] ,
	\end{aligned}
\end{equation} 
and the third fixed point is a stable spiral [Fig.~\ref{dynsys}].

Within this picture, one would expect interfaces between the phases to correspond to heteroclinic orbits that connect these two fixed points~\cite{caussin_emergent_2016}. However, Eq.~\eqref{system1} can support at most one heteroclinic trajectory, which goes from the gaseous fixed point $(q_1^g,m_1^g)$ to the liquid one $(q_1^l,m_1^l)$. This corresponds to the trailing interface. The entire portion of phase space above this heterocline belongs to the basin of attraction of the stable spiral. Indeed, for any choice $V_1$ and $c$, there are no orbits that start from the liquid fixed point $(q_1^l,m_1^l)$ and asymptotically approach the gaseous fixed point $(q_1^g,m_1^g)$  [Fig.~\ref{dynsys}(a)]. This means that the leading interface of the travelling profile is not captured by Eq.~\eqref{system1}.

To resolve this issue, we consider the next order term in the equation for $m_1$: this contribution has a non-perturbative effect on the phase flow. In the region of the portrait with very small $m_1$, it is necessary to account for terms in Eq.~\eqref{system1} at ${\cal O}(\text{Pe}^{3/2})$: one finds $\dot m_1 = \Pe (c-q_1) + {\cal O}(m_1)$ in this region. The effect of this term may be accounted for by a composite equation
\begin{equation}\label{system1b}
\begin{aligned}
	\dot{m}_1 &= -\frac{2\beta}{\varphi_g^2V_1}m_1\left(q_1-\frac{m_1^2}{3\beta}\right)+\text{Pe}\left(c-q_1\right) ,
		\\
	\dot{q_1} &= V_1\left(c-q_1\right) + m_1-\frac{2\beta^2}{\varphi_g^4V_1}m_1^2\left(q_1-\frac{m_1^2}{3\beta }\right) ,
\end{aligned}
\end{equation}
which is valid for small $\Pe$: the term proportional to $\Pe$ is the singular perturbation. Interpreting these equations as a dynamical system, the phase portrait is shown in Fig.~\ref{dynsys}(b). There is now a second heterocline connecting the liquid and vapor fixed points. This effect can be traced back to a change in topology of the nullclines, which is the non-perturbative effect at small Pe. Figure~\ref{dynsys}(b) shows that the basin of attraction of the stable spiral is now contained between the two heteroclines.

Physically, the upper heterocline in Fig.~\ref{dynsys}(b) corresponds to the leading edge of the travelling band. Setting $\Pe=0$ in Eq.~\eqref{system1b} yields \eqref{system1} and this line disappears.  We must instead take $\Pe\to0$, in which case the limiting heterocline passes through the origin [$(q_1,m_1)=(0,0)$] which corresponds to $(\bar\rho,\bar m)=(\varphi_g, 0)$: this is a point on the gaseous spinodal.  The segment of the heterocline with $q_1<0$ connects the origin to the fixed point $(q_1^g,0)$ which corresponds to a point $(\bar\rho,\bar m)=(\rho_g, 0)$ on the gaseous binodal.  This is exactly the scenario found in simulations of the hydrodynamics [Fig.~\ref{lowpe}] where the leading edge of the interface has a segment with $m\approx 0$, between the spinodal and binodal densities.


\section*{References}

\bibliographystyle{iopart-num}
\bibliography{references}

\end{document}